\documentclass[12pt,a4paper,fleqn]{article}
\usepackage{bbm}
\usepackage{amsfonts}
\usepackage[fleqn]{amsmath}
\usepackage{amsmath}
\usepackage{mathrsfs}
\usepackage{amssymb}
\usepackage{latexsym}
\usepackage{stmaryrd}
\usepackage{makeidx}
\usepackage[all]{xy}
\usepackage{CJK}
\usepackage{indentfirst}
\usepackage{tikz}
\usepackage{array}
\usepackage{graphicx}
\usepackage{epsfig}
\usepackage{subfigure}
\usepackage{color}
\usepackage{cite}
\usepackage[percent]{overpic}
\usepackage{overpic}
\usepackage[top=0.5in, bottom=1in, left=0.5in, right=0.5in]{geometry}
\usepackage[colorlinks, linkcolor=red, anchorcolor=blue, citecolor=blue]{hyperref}
\newcommand{\ds}{\displaystyle}
\date{}

\title{\Large \textbf{SUSY SU(5) $\times S_{4}$ GUT Flavor Model for Fermion Masses and Mixings with Adjoint, Large $\theta^{PMNS}_{13}$}}
\author{
Ya Zhao  $^{1, }$ \footnote{\emph{E-mail address}: zhaoya@mail.ustc.edu.cn} ,
Peng-Fei Zhang  $^{1, }$ \footnote{\emph{E-mail address}: zhpf@ustc.edu.cn} \\
\bigskip
\\
{ $^{1}$ {\footnotesize
 \it Department of Modern Physics, University of Science and Technology  of China, Hefei, Anhui 230026, China.}}}
\begin{document}
\normalsize
\maketitle
\begin{abstract}
  We propose an $S_{4}$ flavor model based on supersymmetric (SUSY) SU(5) GUT. The first and third generations of \textbf{10} dimensional
  representations in SU(5) are all assigned to be $1_{1}$ of $S_{4}$. The second generation of \textbf{10} is to be $1_{2}$ of $S_{4}$.
  Right-handed neutrinos of singlet \textbf{1} and three generations of $\bar{\textbf{5}}$ are all assigned to be $3_{1}$
  of $S_{4}$. The VEVs of two sets of flavon fields are allowed a moderate hierarchy, that is
  $\langle\Phi^{\nu}\rangle \sim \lambda_{c}\langle\Phi^{e}\rangle$. Tri-Bimaximal (TBM) mixing can be produced at both leading order (LO)
  and next to next to leading order (NNLO) in neutrino sector. All the masses of up-type quarks are obtained at LO. We also get the
  bottom-tau unification $m_{\tau}=m_{b}$ and the popular Georgi-Jarlskog relation $m_{\mu}=3m_{s}$ as well as a new
  mass relation $m_{e}=\frac{8}{27}m_{d}$ in which the novel Clebsch-Gordan (CG) factor arises from the adjoint field
  $H_{24}$. The GUT relation leads to a sizable mixing angle $\theta^{e}_{12} \sim \theta_{c}$ and the correct quark mixing matrix $V_{CKM}$ can
  also be realised in the model. The resulting CKM-like mixing matrix of charged leptons modifies the vanishing $\theta^{\nu}_{13}$ in TBM
  mixing to a large $\theta^{PMNS}_{13}\simeq\theta_{c}/\sqrt{2}$, in excellent agreement with experimental results. A Dirac CP violation
  phase $\phi_{12}\simeq\pm\pi/2$ is required to make the deviation from $\theta^{\nu}_{12}$ small. We also present some
  phenomenological numerical results predicted by the model.
\end{abstract}
\section{Introduction}\label{S1}
In the Standard Model (SM) of particle physics the charged fermions, quarks and charged leptons, are massive fermions, while neutrinos are
massless in SM. However the deficit of the observed neutrinos with respect to the theoretical predicted ones leads to the two famous puzzles in
neutrino physics, i.e., the longstanding solar neutrino puzzle~\cite{Davis/1968SNP} and the atmospherical neutrino
anomaly~\cite{Haines/1986ANP, Hirata/1988ANP} before 1998. The puzzles can be explained through the  neutrino oscillation mechanism,
which indicate that the neutrinos are also massive and lepton flavors are mixed. The discovery of neutrino
oscillations~\cite{Fukuda/1998SK,Ahmad/2002SNO} convinced people that neutrinos have tiny masses.
Meanwhile the seesaw mechanism~\cite{Minkowski/1977ss,Yanagida/1979ss,Mohapatra/1980ss,Gell-Mann/1979sug} seems to be a
graceful solution to answer why neutrino masses are small. Nevertheless, the flavor mixing pattern with the observed mixing angles can not be
explained through seesaw mechanism. Solar and atmospherical neutrino oscillation experiments have measured leptonic mixing angles with
great accuracy. The resulting lepton mixing Pontecorvo-Maki-Nakagawa-Sakata matrix $U_{PMNS}$~\cite{Pon/1968, MNS/1962} can be well compatible
with the simple Tri-Bimaximal (TBM) mixing pattern, introduced by Harrison, Perkins and Scott~\cite{HPS/2002}:
\begin{equation}\label{eq:TBM}
\qquad    U_{TB}=\left(
             \begin{array}{ccc}
               \sqrt{\frac{2}{3}} & \frac{1}{\sqrt{3}} & 0 \\
               -\frac{1}{\sqrt{6}} & \frac{1}{\sqrt{3}} & -\frac{1}{\sqrt{2}} \\
               -\frac{1}{\sqrt{6}} & \frac{1}{\sqrt{3}} & \frac{1}{\sqrt{2}} \\
             \end{array}
           \right)
\end{equation}
which predicts that $\theta^{\nu}_{23}=\frac{\pi}{4}$ and $\theta^{\nu}_{12}=\arcsin\frac{1}{\sqrt{3}}$, but $\theta^{\nu}_{13}=0$. The two nonzero
leptonic mixing angles $\theta^{\nu}_{12}$ and $\theta^{\nu}_{23}$ are predicted to be rather large by contrast with the quark mixing angles, which are
known to be very small~\cite{PDG/2012}. Besides the Tri-Bimaximal mixing pattern \textit{ansatz}, similar simple mixing patterns with vanishing
$\theta^{\nu}_{13}$ were proposed, such as Bimaximal (BM)~\cite{BM/1998,Alta/2009a, Melo/2011a}, Golden-Ratio (GR)~\cite{GR/2007,
Everett/2009,Rode/2009a,Adul/2009d,Ding/2012b} and Democratic~\cite{DE/1996} mixing patterns.
The simple patterns suggest some kind of underlying non-Abelian discrete flavor symmetry $G_{f}$ would exist in the lepton sector at least.
Indeed the mixing patterns have been paid a lot of attention in the flavor model buidlding community.
For the flavor models based on the typical discrete symmetries, please see Refs.~\cite{Alta/2010,Ish/2010a,King/rev} for a review.
The models based on continuous groups have been proposed~\cite{King/2005,deMe/2006,deMe/2007,King/2006,Antusch/2008,Berger/2010,Adul/2009}.
By adding higher order corrections, most of the models can give rise to a non-zero reactor angle $\theta_{13} \sim \mathcal {O}(\lambda^{2}_{c})$,
with $\lambda_{c} = \sin\theta_{c}\simeq 0.22$ being Wolfenstein parameter~\cite{Wolfen/1983},
where $\theta_{c}$ is the Cabibbo angle. The resulting small $\theta_{13}$ was within the range of global 
fits~\cite{Fogli/2008} before the determinate large $\theta_{13}$ Daya Bay~\cite{dyb/2013} neutrino experiment measured. The deviations from the
TBM values of $\theta_{12}$ and $\theta_{23}$ are also at most $\mathcal {O}(\lambda^{2}_{c})$ when subleading effects
are included, which is in agreement at 3$\sigma$ error range with the experimental data or say global fits.

The Daya Bay Collaboration~\cite{dyb/2013} now has confirmed a larger $\theta_{13}$ with a significance of 7.7 (the first is 5.2)
standard deviations from the reactor $\overline{\nu}_{e}\rightarrow\overline{\nu}_{e}$ oscillations. The best-fit result in $1\sigma$ range is
\begin{equation}\label{eq:dyb}
\qquad    \sin^{2}2\theta_{13}=0.089\pm0.010(stat)\pm0.005(syst),
\end{equation}
which is equivalent to $\theta_{13}\simeq8.7^{\circ} \pm 0.8^{\circ}$. And the RENO~\cite{RENO/2012} also reported that
\begin{equation}\label{eq:reno}
\qquad    \sin^{2}2\theta_{13}=0.113\pm0.013(stat)\pm0.019(syst).
\end{equation}
The updated Daya Bay~\cite{HuBZ/2015dyb} data is measured to remarkable accuracy: $\sin^{2}2\theta_{13}=0.084\pm0.005$
or $\theta_{13}\approx8.4^{\circ}\pm0.2^{\circ}$. Even before the Daya Bay result, however, there have emerged direct evidence of large
$\theta_{13}$ from T2K~\cite{T2K/2011}, MINOS~\cite{MINOS/2011a} and Double Chooz \cite{DC/2012}.
The accurate nonzero reactor angle $\theta_{13}$ implies Tri-Bimaximal mixing pattern and
else would be ruled out. However there still exists the possibility of maintaining the mixing angles $\theta^{\nu}_{12}=\arcsin\frac{1}{\sqrt{3}}$
and $\theta^{\nu}_{23}=\frac{\pi}{4}$ which TBM predicted as the leading order (LO) result of a model. The phenomenological favored value
$\theta_{13}\approx\lambda_{c}/\sqrt{2}$, although now is intension with the updated data, can be derived from some proper corrections. In a general
scheme consideration for obtaining the large $\theta_{13}$ from zero, the most popular and well motivated correction to TBM mixing is the
contributions from charged lepton mixing, especially inducing a sizable $\theta^{e}_{12}\sim \theta_{c}$ is viable in the Grand Unified Theories
(GUT) flavor models, see~\cite{Antusch/2011,Marzo/2011} as example. The gauge symmetry groups are usually chose as SU(5), SO(10) or Pati-Salam
context SU(4)$_{C}\times$ SU(2)$_{L} \times$ SU(2)$_{R}$. The simplest GUT gauge symmetry group is SU(5)~\cite{GeGl/1974}, in which matter fields
of standard model are assigned to be $\bar{\textbf{5}}$ and $\textbf{10}$ dimensional representations. In a generical GUT scheme the
down-type quark and charged lepton Yukawa matrices are the crucial factor to produce GUT relations that connect $\theta_{13}$ with Cabibbo angle,
see~\cite{Antusch/2013a,Antusch/2013b}. We remark that a series of models based on discrete flavor symmetry group together with a GUT gauge group have
been proposed, for example the SU(5) $\times A_{4}$~\cite{Alta/2008,Cial/2009,Burr/2010},
SU(5)$\times S_{4}$~\cite{Ish/2009a,Ish/2009b,Hagedorn/2010,Ding/2011,Melo/2011a,Hagedorn/2012}
and SU(5)$\times T'$~\cite{ChenMC/2007}, SO(10)$\times A_{4}$~\cite{Morisi/2007,Bazz/2008,Albaid/2009}, SO(10)$\times
S_{4}$~\cite{LeeDG/1994,Mohap/2004,CaiY/2006,Dutta/2010,Patel/2011,BhDev/2011,BhDev/2012}, SO(10)$\times PSL_{2}(7)$~\cite{King/2009,King/2010}
and SO(10)$\times \Delta_{27}$~\cite{Bazz/2009c}. Most of the GUT flavor models also generically give rise to
$\theta_{13}\sim \mathcal{O}(\lambda^{2}_{c})$, while only few models based on different ansatz, such as Bimaximal~\cite{Alta/2009a,Melo/2011a},
or empirical relation such as Quark-Lepton Complementarity (QLC)~\cite{Patel/2011} may lead to sizable $\theta_{13}\sim \mathcal{O}(\lambda_{c})$.
The other way to achieve the sizable $\theta_{13}$ is the introduction of non-singlets such as the SU(5) adjoint fields $\textbf{24}$, which split
the heavy messenger masses and give rise to new novel GUT Yukawa coupling ratios of quark-lepton. For the realistic GUT flavor models based on the
mechanism one can refer to, such as~\cite{Antusch/2013gA4,Gehrlein/2015gA5,Bjorkeroth/2015gA4}.

In this paper we propose a SUSY SU(5) GUT flavor model, with $S_{4}\times Z_{4} \times Z_{6}\times Z_{5} \times Z_{2}$ as flavor symmetry groups.
The flavor symmetry $S_{4}$ can be spontaneously broken by Vacuum Expectation Values (VEV) of flavon fields in $\Phi$ which is divided into
$\Phi^{e}$ in charged fermion sector and $\Phi^{\nu}$ in neutrino sector. The assumption we adopted is similar with Ref.\cite{Lin/2010},
of which the VEVs of $\Phi^{e}$ and $\Phi^{\nu}$ allow a moderate hierarchy: $\langle\Phi^{\nu}\rangle \sim \lambda_{c}\langle\Phi^{e}\rangle$.
The dynamical tricky assumption makes the $\theta^{PMNS}_{13}$ around $\mathcal{O}(\lambda_{c})$ possible. The neutrino masses are simply
generated through type-I see-saw mechanism. Tri-Bimaximal mixing pattern is produced exactly at LO, and even still holds exactly at next to
next to leading order (NNLO) in neutrino sector, which is a salient feature of our model. The charged fermion mass hierarchies are controlled
by spontaneously broken of the flavor symmetry without introducing Froggatt-Nielsen mechanism~\cite{FN/1979}. Both up- and down-type
quarks obtain their masses with proper order of magnitude and the correct quark mixing matrix $V_{CKM}$ can be realised in the model.

The masses of charged leptons are similar with those of down-type quarks regardless of the different group-theoretical Clebsch-Gordan 
(CG) coefficients. Due to the introduction of an adjoint field $H_{24}$, the resulting novel CG factors lead to a new mass ratio between electron 
and down quark, namely $m_{e}/m_{d}=8/27$, which is a phenomenological favored result. The famous Georgi-Jarlskog relation $m_{\mu}=3m_{s}$ 
and bottom-tau unification $m_{\tau}=m_{b}$ are also maintained in the model. The mixing angle $\theta^{e}_{12}\sim \theta_{c}$ is also achieved by the
specific GUT-scale relation between the angle and mass ratio $m_{e}/m_{\mu}$~\cite{Antusch/2011}. Finally the CKM-like mixing matrix of charged
leptons would modify the vanishing $\theta^{\nu}_{13}$ in TBM mixing to a large $\theta^{PMNS}_{13}\simeq \theta_{c}/\sqrt{2}$, in excellent
agreement with experimental determinations.

The paper is organized as follows. In Section~\ref{S2} we discuss the basic strategic considerations for obtaining the
large $\theta^{e}_{12}\sim\lambda_{c}$, including the desired Yukawa textures, the hierarchy assumption on flavon VEVs and
the role of the adjoint $H_{24}$. In Section~\ref{S3} we introduce all matter fields and flavons, and the predictions
for the fermions masses and mixings at LO are presented. In Section~\ref{S4} the vacuum alignment are justified by minimizing the potential.
Section~\ref{S5} is devoted to the subleading corrections to the VEVs of the flavons, the LO masses and mixings of fermions. In Section~\ref{S6} 
we show a bit of phenomenology of the model predicted in numerical results. Section~\ref{S7} is our conclusion.
\section{The strategy and assumptions}\label{S2}
In a large class of flavor models that give arise to TBM and else mixing patterns with or without GUT context, the angle $\theta_{13}$ is usually 
about $\mathcal{O}(\lambda^{2}_{c})\sim 3^{\circ}$ by taking subleading corrections into account. In order to obtain the sizable mixing angle 
$\theta_{13}\sim\mathcal{O}(\lambda_{c})$, the most popular and well-motivated correctional approach can be provided
by the large charged lepton mixing contributions. The fully lepton mixing matrix $U_{PMNS}=V^{\ell\dagger}_{L}U_{\nu}$
in which $U_{\nu}$ is usually taken TBM, BM or GR as first order approximation, while $V^{\ell}_{L}$ is not uniquely determined.
The general model-independent studies on the deviations from TBM mixing with the contributions
of $V^{\ell}_{L}$ have been proposed, see Refs.~\cite{Plent/2005,Hoch/2007,HeXG/2007,Rode/2009b,HeXG/2011,Kitab/2012} as examples.

One of the aims in the work is to generate a large mixing $\theta^{e}_{12}$ in charged lepton sector. In the context of unified
theory the Yukawa matrices of charged leptons and down quarks are unified in single joint operators, which provide a possible approach
to generating a larger mixing angle $\theta^{e}_{12}\simeq\lambda_{c}$. Nevertheless, the resulting mixing angle and masses should
satisfy some specific GUT-scale relations in order to fit the realistic phenomenological constrains. However the traditional
unified operators which give rise to some GUT-scale mass relations, such as the popular Georgi-Jarlskog (GJ) relations~\cite{GJ/1979}
$m_{\mu}=3m_{s}$ and $m_{e}\simeq\frac{1}{3}m_{d}$ will lead to $\theta^{e}_{12}\simeq\lambda_{c}/3$. As consequence the GJ factor of 3 leads
to $\theta_{13}\simeq\lambda_{c}/3\sqrt{2}$ in a large class of GUT flavor models, which now contradicts with experimental data.
For achieving $\theta_{13}\simeq\lambda_{c}/\sqrt{2}$ in GUT flavor models, the basic strategic considerations has been suggested
in~\cite{Antusch/2011,King/2012b,Antusch/2013a,Antusch/2013b}. A different strategy without GUT can be seen in~\cite{Alta/2012a}.
Note that the Yukawa matrices we mentioned are taken to be equivalent to the mass matrices, or say $m_{ij}=y_{ij}v$
with the value of Higgs VEV $v$ being fixed at the electroweak scale $\Lambda_{EW}$. Without loss of generality we
consider the upper $2\times2$ part of down quarks Yukawa matrices with a vanishing 11-entry for simplicity, the desired Yukawa matrices
that lead to $\theta^{d,e}_{12}\simeq\lambda_{c}$ are generically of the form
\begin{equation}\label{eq:yukawaDe}
 \qquad Y_{D} \propto \left(
                   \begin{array}{ccc}
                     0 & a\\
                     b & c \\
                   \end{array}
                 \right)\Longrightarrow \quad
 Y_{e} \propto \left(
                   \begin{array}{ccc}
                     0 & c_{a}a\\
                     c_{b}b & c_{c}c \\
                   \end{array}
                 \right)^{T}= \left(
                   \begin{array}{cc}
                     0 & c_{b}b\\
                     c_{a}a & c_{c}c \\
                   \end{array}
                 \right)
\end{equation}
in which parameters $a, b$ and $c$ are suitable complex numbers that can give rational magnitude order of the quark masses $m_{d}\simeq|ab/c|$
and $m_{s}\simeq|c|$ and the desired mixing angle $\theta^{d}_{12}\simeq|b/c|\simeq \lambda_{c}$, thus the Cabibbo angle can be derived.
Similarly the charged lepton masses $m_{e}=|\frac{c_{a}ac_{b}b}{c_{c}c}|$ and $m_{\mu}\simeq c_{c}c$ and the mixing angle
$\theta^{e}_{12}\simeq|\frac{c_{a}a}{c_{c}c}|$, where the CG factors $c_{a,b,c}$ are uniquely determined by SU(5) contractions.
The mixing angle $\theta^{e}_{12}$ is, however, constrained by the GUT-scale relation between mixing angle and the mass ratio~\cite{Antusch/2011}:
$\theta^{e}_{12}=|\frac{c_{c}}{c_{b}}|\frac{m_{e}}{m_{\mu}}\frac{1}{\theta^{d}_{12}}$. The mixing angle
$\theta^{e}_{12}\simeq \lambda_{c}$ can be achieved only when the CG ratio $|\frac{c_{c}}{c_{b}}|\sim \mathcal{O}(2/\lambda_{c})$ is
satisfied since the order of the mass ratio $\frac{m_{e}}{m_{\mu}}\sim \lambda^{3}_{c}/2$ and $\theta^{d}_{12}\sim \lambda_{c}$.

For satisfying the ratio the CG factors in the Yukawa matrix $Y_{e}$ should be chose properly such as the novel
ones in~\cite{Antusch/2011,Antusch/2014GUT}. To be specific the structures of $Y_{e}$ are given as following
\begin{equation}\label{eq:yukawa}
 \qquad  Y_{e} \propto
 \left(
                   \begin{array}{cc}
                     0 & -\frac{1}{2}b \\
                     6a & 6c \\
                   \end{array}
                 \right) \qquad \text{or}\qquad
 \left(
                   \begin{array}{ccc}
                     0 & \frac{4}{9}b \\
                     \frac{9}{2}a & \frac{9}{2}c \\
                   \end{array}
                 \right),
\end{equation}
which predict $|\frac{c_{c}}{c_{b}}|=12$ or $|\frac{c_{c}}{c_{b}}|= 81/8$. Both Yukawa textures satisfy the ratio
$|\frac{c_{c}}{c_{b}}|\sim \mathcal{O}(2/\lambda_{c})$ and also require $a/c\simeq\lambda_{c}$. The above structures demand
the introduction of non-singlet fields, such as adjoint fields $\textbf{24}$ in either the numerator (first $Y_{e}$) or the denominator
(second $Y_{e}$) of the effective operators. Only in the methods can we obtain new predictions of Yukawa coupling ratios.
For such realistic models one can refer to~\cite{Antusch/2013gA4,Gehrlein/2015gA5,Bjorkeroth/2015gA4}. However we find that there exists
another possible structure of $Y_{e}$ which preserves the Georgi-Jarlskog relation and
the ratio $|\frac{c_{c}}{c_{b}}|= 81/8$ still holds. The texture of $Y_{e}$ is the metamorphosis of the second $Y_{e}$ in Eq.~(\ref{eq:yukawa})
\begin{equation}\label{eq:yukawa2}
  \qquad Y_{e} \propto
 \left(
                   \begin{array}{cc}
                     0 & -\frac{8}{27}b \\
                     -3a & -3c \\
                   \end{array}
                 \right).
\end{equation}

It is an important goal to obtain above $Y_{e}$ in the model. The CG factor $-3$ arises from the conventional GJ Higgs $H_{\overline{45}}$, and the
novel CG factor $-\frac{8}{27}$ is due to the VEV of the adjoint $H_{24}$ appears in the denominator of effective operators. The non-singlet
fields are essential to make the masses (size around GUT scale) of the components of the messenger split by
CG coefficients~\cite{Antusch/2014GUT}. Finally the CG factors enter inversely in the desired Yukawa matrix elements and the new predictions arise.

Another difficulty in generating the desired Yukawa structures and hence the mixings is the vacuum alignments which arise from the spontaneously
broken flavor symmetry. Denoting the general scalar flavon fields with $\Phi$, $S_{4}$ can be spontaneously broken by the VEVs of flavons in $\Phi$
which is divided into two sets: $\Phi^{e}=\{\varphi,\eta,\chi,\sigma,\vartheta,\xi,\rho\}$ in charged fermion sector 
and $\Phi^{\nu}=\{\phi,\Delta,\zeta\}$ in neutrino sector. We also assume a moderate hierarchy between the VEVs of $\Phi^{e}$ and $\Phi^{\nu}$:
$\langle\Phi^{\nu}\rangle \sim \lambda_{c}\langle\Phi^{e}\rangle$, to induce $a/c\simeq\lambda_{c}$. For sake of convenience two small expansion
parameters $\epsilon$ and $\delta$ are introduced as
\begin{equation}\label{eq:assumption}
\qquad  \frac{\langle\Phi^{e}\rangle}{\Lambda}\sim\epsilon\sim\lambda^{2}_{c},
\quad \frac{\langle\Phi^{\nu}\rangle}{\Lambda}\sim\delta\sim\lambda^{3}_{c}.
\end{equation}
The assumption provides a possibility to generate the desired relation between (21) and (22) elements in Eq.~\eqref{eq:yukawa}. In fact the angle
$\theta^{e}_{12}\sim\lambda_{c}$ in the model is produced by the ratio $\langle\Phi^{\nu}\rangle/\langle\Phi^{e}\rangle\sim\lambda_{c}$. The
specific dynamical tricks are in principle allowed by ``separated" scalar potential, which is guaranteed by the auxiliary Abelian flavor symmetry
$\mathcal{G}_{A}$. The $\mathcal{G}_{A}$ separates the scalar potential of the flavons in $\Phi^{e}$ and $\Phi^{\nu}$ generically as follows
\begin{equation}\label{eq:Vsep}
\qquad  V(\Phi^{e},\Phi^{\nu})=V_{\nu}(\Phi^{\nu})|_{\textrm{LO}}+V_{e}(\Phi^{e})|_{\textrm{LO}}+V(\Phi^{\nu}, \Phi^{e})|_{\textrm{sub}}
\end{equation}
where $V(\Phi^{\nu}, \Phi^{e})|_{\textrm{sub}}$ is the subleading scalar potential, at least at NLO, and is usually called
``partially" separated.  At LO the two sectors are naturally separated, but it is not the case for subleading corrections at NLO and/or NNLO.
The scalar potential is ``fully" separated while
$V(\Phi^{\nu}, \Phi^{e})|_{\textrm{sub}}=V(\Phi^{e})|_{\textrm{sub}}$ or $V(\Phi^{\nu}, \Phi^{e})|_{\textrm{sub}}=V(\Phi^{\nu})|_{\textrm{sub}}$,
which makes a hierarchy between $\langle\Phi^{e}\rangle$ and $\langle\Phi^{\nu}\rangle$ possible, see Lin's work in Ref.\cite{Lin/2010}.
In our model the auxiliary flavor group is chosen as $\mathcal{G}_{A}=Z_{4}\times Z_{6}\times Z_{5}\times Z_{2}$.

However in the present GUT flavor model the ``fully" separated scalar potential is not exactly the same as that in Lin's proposal.
The subleading scalar potential is separated as
\begin{equation}\label{eq:Vsub}
  \qquad V(\Phi^{\nu},\Phi^{e})|_{\textrm{sub}}=V(\Phi^{\nu})|_{\textrm{NNLO}}
    +V(\Phi^{e},\Phi^{\nu})|_{\textrm{NLO}}+\cdots
\end{equation}
where the dots stand for higher order subleading terms. The scalar potential of $\Phi^{\nu}$ is separated at both LO and NNLO, thus the magnitude of
$\langle\Phi^{\nu}\rangle/\Lambda$ is not necessarily the same as that of $\langle\Phi^{e}\rangle/\Lambda$. Actually it is feasible to build a model
for TBM based on the $S_{4}$ symmetry with the allowed hierarchy in Eq.~\eqref{eq:assumption}. The hierarchy assumption not only gives arise to the
large $\theta^{e}_{12}\sim\lambda_{c}$, but also produces the correct up quark mass at NLO. The mass hierarchies of up-type quarks are roughly as
$m_{u}:m_{c}:m_{t}=\lambda^{8}_{c}:\lambda^{4}_{c}:1$, we note that $\lambda^{8}_{c}$ can be not only given by $\epsilon^{4}$, but also by
$\epsilon\delta^{2}$. Indeed the mass of up quark is given by the flavon combinations of order $\epsilon\delta^{2}$ at leading order operators.
To a certain extent the hierarchical VEVs of $\Phi^{e}$ and $\Phi^{\nu}$ are even necessary in the present model.

\section{The Construction of the Model}\label{S3}
In this section we introduce the SUSY SU(5)$\times S_{4}$ GUT flavor model with the auxiliary Abelian $Z_{4}\times Z_{6}\times Z_{5}\times Z_{2}$ 
shaping symmetries. The flavor symmetry group $S_{4}$ is the permutation group of four objects, as well as the invariance group of octahedron and 
cube. It has 24 elements, which can be generated by two basic permutations \textit{S} and \textit{T} as the generators :
\begin{equation}\label{eq:ST}
    \qquad S^{4}=T^{3}=(ST^{2})^{2}=1
\end{equation}
In group theory the generic permutation can be expressed by $(1,2,3,4)\rightarrow(n_{1},n_{2},n_{3},n_{4})\equiv(n_{1}n_{2}n_{3}n_{4})$. The two basic
permutations \textit{S} = (2341) and \textit{T} = (2314) is used to generate the elements of $S_{4}$.
The group has five inequivalent irreducible representations: two three-dimensional representations ${3}_{1}$ and $3_{2}$, one 2-dimensional 2,
and two one-dimensional $1_{1}$ and $1_{2}$ representations. The multiplication rules are presented as follows:
\begin{eqnarray}\label{eq:rule}
    & & 1_{1}\otimes r= r \otimes 1_{1} = r, \quad 1_{2} \otimes 1_{2}=1_{1}, \quad  1_{2}\otimes 2=2,
    \quad 1_{2}\otimes 3_{1}=3_{2}, \quad 1_{2}\otimes 3_{2}=3_{1}, \nonumber\\
    & & 2\otimes 2=1_{1}\oplus1_{2}\oplus2, \quad 2\otimes3_{1}=3_{1}\oplus3_{2}, \quad 2\otimes 3_{2}=3_{1}\oplus3_{2}, \nonumber\\
    & & 3_{1}\otimes3_{1}=3_{2}\otimes3_{2}=1_{1}\oplus2\oplus3_{1}\oplus3_{2}, \quad 3_{1}\otimes3_{2}=1_{2}\oplus2\oplus3_{1}\oplus3_{2}
\end{eqnarray}
The detailed irreducible representation matrices and the matrices of generators \textit{S}, \textit{T} are presented in \hyperref[A]{APPENDIX A}.
In the model matter fields, Higgs and flavon fields are assigned to be different representations of gauge group SU(5). All matter fields in SU(5) are
unified into $\bar{\textbf{5}}$ and $\textbf{10}$ dimensional representations, denoted by $F$ and $T_{1, 2, 3}$, respectively. The Higgs fields
include the SU(5) $\textbf{5}$, $\bar{\textbf{5}}$, $\textbf{45}$ and $\overline{\textbf{45}}$-dimensional representations. The only one adjoint
$\textbf{24}$-dimensional field, $H_{24}$, is the key to obtain the desired CG factors. The right handed neutrinos $N^{c}$ and all flavon fields 
in $\Phi$ are SU(5) gauge singlets $\textbf{1}$. All the fields are also assigned to be different representations of the flavor symmetry group 
$S_{4}$ and the auxiliary shaping symmetries $Z_{N}$. The first and third generations of \textbf{10} dimensional representations $T_{1}$ and 
$T_{3}$ are all assigned to be $1_{1}$ of $S_{4}$ singlet. The second generation of \textbf{10} dimensional representation $T_{2}$ is assigned 
to be $1_{2}$ of $S_{4}$. Right-handed neutrinos $N^{c}$ and three generations of $\bar{\textbf{5}}$ $F$ are all assigned to be $3_{1}$ of $S_{4}$. 
The Higgs $H_{5,\bar{5},\overline{45}}$ and the adjoint $H_{24}$ are all assigned to be $1_{1}$ of $S_{4}$, while $H_{45}$ is to be $1_{2}$ of 
$S_{4}$. The flavon fields in $\Phi$, which include all possible $S_{4}$ representations, are introduced to break the $S_{4}$ flavor 
symmetry spontaneously. To be specific, the left-handed down-type quarks in three colors and doublet leptons are collected in SU(5) 
representation $\bar{\textbf{5}}$ as
\begin{equation}\label{eq:F5}
\qquad  F = (d^{c}_{R} \quad d^{c}_{B} \quad d^{c}_{G} \quad e \quad -\nu)
\end{equation}
whereas the representation \textbf{10} contains $SU(2)_{L}$ doublet quarks as well as up-type quark singlet and charged leptons
\begin{equation}\label{eq:T10}
\qquad  T_{i}=\frac{1}{\sqrt{2}}\left(
                            \begin{array}{ccccc}
                              0 & -u^{c}_{G} & u^{c}_{B} & -u_{R} & -d_{R} \\
                              u^{c}_{G} & 0 & -u^{c}_{R} & -u_{B} & -d_{B} \\
                              -u^{c}_{B} & u^{c}_{R} & 0 & -u_{G} & -d_{G} \\
                              u_{R} & u_{B} & u_{G} & 0 & -e^{c} \\
                              d_{R} & d_{B} & d_{G} & e^{c} & 0 \\
                            \end{array}
                          \right)_{i}
\end{equation}
where $i=1,2,3$ indicates the fermions family indices of standard model (SM), and $R, B, G$ stand for the color indices. The matter fields and Higgs in
the SU(5)$\times S_{4}\times Z_{4}\times Z_{6}\times Z_{5}\times Z_{2}$ model, with their transformation properties under the flavor symmetry group,
are listed in \hyperref[ta:tab1]{Table 1}. The flavon fields and the additional gauge singlets $\Phi_{0}$, the driving fields $\varphi_{0}$,
$\eta_{0}$, $\chi_{0}$, $\sigma_{0}$, $\xi_{0}$ and $\phi_{0}$, $\Delta_{0}$, $\zeta_{0}$, are listed in \hyperref[ta:tab2]{Table 2}. We also
introduce a global $U(1)_{R}$ continuous symmetry which meant a R-parity discrete subgroup. The driving fields carry +2 $U(1)_{R}$ charge,
which made them linearly appear in the superpotential. Matter fields and heavy right-handed neutrinos are charged with +1 $U(1)_{R}$ charge,
while all Higgs fields and flavons are uncharged.

\begin{table}[!htbp]\footnotesize
  \centering
  \caption{Transformation properties of the matter fields and Higgs fields in the model, where $\omega=e^{i\pi/3}$ for $Z_{6}$
  group and $\omega=e^{i2\pi/5}$ for $Z_{5}$ group.}\label{ta:tab1}
  \begin{tabular}{|ccccccccccc|}
  \hline
  \hline
  Field  & $T_{3}$ & $T_{2}$ & $T_{1}$ & $F$ & $N^{c}$ & $H_{5}$ & $H_{\bar{5}}$ & $H_{45}$ & $H_{\overline{45}}$ & $H_{24}$\\
  \hline
  SU(5) & \textbf{10} & \textbf{10} & \textbf{10}& $\bar{\textbf{5}}$ & \textbf{1} & \textbf{5} & $\bar{\textbf{5}}$ & \textbf{45} &
  $\overline{\textbf{45}}$ & \textbf{24} \\
   $S_{4}$ & $1_{1}$  & $1_{2}$ & $1_{1}$ & $3_{1}$ & $3_{1}$ &$1_{1}$ &$1_{1}$ &$1_{2}$ &$1_{1}$ & $1_{1}$ \\
  $Z_{4}$ & 1 & 1 & $-1$ & 1 & 1 & 1 & $-1$ & $i$ & $i$ & $-1$ \\
  $Z_{6}$ & $\omega$ & $-\omega$ & 1 & $\omega^{2}$ & 1 & $-\omega$ & $\omega$ & $-\omega$ & $-\omega^{2}$ & $\omega$ \\
  $Z_{5}$ & 1  & $\omega$ & $\omega^{2}$ & 1 & 1 & 1 & $\omega^{4}$ & 1 & $\omega^{3}$ & 1  \\
  $Z_{2}$ & 1  & 1 & $-1$ & 1 & $-1$ & 1 & $-1$ & $-1$ & $-1$ & 1  \\
  $U(1)_{R}$ & 1 & 1 & 1 & 1 & 1 & 0 & 0 & 0 & 0 & 0 \\
  \hline
  \hline
  \end{tabular}
\end{table}

\begin{table}[!htbp]\footnotesize
  \centering
  \caption{Transformation properties of flavons and driving fields in the model}\label{ta:tab2}
  \resizebox{\textwidth}{!}{
  \begin{tabular}{|ccccccccccc|cccccccc|}
  \hline
  \hline
  Field & $\varphi$ & $\eta$ & $\chi$ & $\sigma$ & $\vartheta$ & $\xi$ &$\rho$ &
   $\phi$ & $\Delta$ & $\zeta$ & $\varphi_{0}$ & $\eta_{0}$ & $\chi_{0}$ & $\sigma_{0}$ & $\xi_{0}$
   & $\phi_{0}$ & $\Delta_{0}$ & $\zeta_{0}$ \\
  \hline
  $S_{4}$ & $3_{1}$ &  2 & $3_{2}$ & $1_{1}$ & $1_{1}$ & $3_{1}$ & $1_{1}$ & $3_{1}$ & 2 & $1_{2}$ &
  $3_{1}$ & $1_{1}$ & $3_{2}$¡¡& $1_{1}$ & $3_{1}$ & $3_{2}$ & 2 & $1_{1}$ \\
  $Z_{4}$ & $-1$ & $-1$ &  1 & $-i$ & $i$ & $i$ & $i$ &1 & 1 & $-1$ &
  1 & 1 & 1 & $-1$ & $-1$ & 1 & 1 & 1 \\
  $Z_{6}$& $\omega^{2}$ & $\omega^{2}$ & $-\omega$ & $-1$ & $-1$ & $\omega$ & $\omega$ & 1 & 1 & 1 &
  $\omega^{2}$ & $\omega^{2}$ & $\omega^{2}$ & 1 & $-\omega$ & 1 & 1 & 1 \\
  $Z_{5}$ & $\omega$ & $\omega$ & $\omega^{2}$ & $\omega^{4}$ & $\omega^{4}$ & $\omega$ & $\omega$ & 1 & 1 & 1 & $\omega^{3}$ &
  $\omega^{3}$ & $\omega^{3}$ & $\omega^{2}$ & $\omega^{3}$ & 1 & 1 & 1  \\
  $Z_{2}$ & $-1$ & $-1$ & 1 & $-1$ & $-1$ & $-1$ & $-1$ & $-1$ & $-1$ & 1 & 1 & 1 & 1 & 1 & 1 & 1 & 1 & 1  \\
  $U(1)_{R}$ & 0 & 0 & 0 & 0 & 0 & 0 & 0 & 0 & 0 & 0 & 2 & 2 & 2 & 2 & 2 & 2 & 2 & 2 \\
  \hline
  \hline
\end{tabular}}
\end{table}

In the following we shall present the model in detail. The masses and mixings of fermions arise from the spontaneously flavor symmetry breaking by
the flavon fields acquiring the VEVs. The alignment directions of the vacua are crucial to generate the observed mass hierarchies and mixings.
For the time being the VEVs of the scalar components of the flavon fields are justified as the natural solutions of the scalar potential
in Section~\ref{S4}, which indicate the structures as following
\begin{eqnarray}\label{eq:VFE}
  & & \langle\varphi\rangle=v_{\varphi}\left(\begin{array}{c}
                                           0 \\
                                           1 \\
                                           0 \\
                                         \end{array}\right),
  \quad \langle\eta\rangle=v_{\eta}\left(\begin{array}{c}
                                     0 \\
                                     1
                                   \end{array}\right),
  \quad \langle\chi\rangle=v_{\chi}\left(\begin{array}{c}
                                           0 \\
                                           0 \\
                                           1 \\
                                         \end{array}\right),
  \quad \langle\xi\rangle=v_{\xi}\left(\begin{array}{c}
                                           1 \\
                                           0 \\
                                           0 \\
                                         \end{array}\right),\nonumber\\
& & \langle\rho\rangle=v_{\rho},\quad\langle\sigma\rangle=v_{\sigma}, \quad \langle\vartheta\rangle=v_{\vartheta},
\end{eqnarray}
for flavons in $\Phi^{e}$ sector and
\begin{eqnarray}\label{eq:VFN}
& & \quad \langle\phi\rangle=v_{\phi}\left(\begin{array}{c}
                                           1 \\
                                           1 \\
                                           1 \\
                                         \end{array}\right),
  \quad \langle\Delta\rangle=v_{\Delta}\left(\begin{array}{c}
                                     1 \\
                                     1
                                   \end{array}\right),
  \quad \langle\zeta\rangle=v_{\zeta}
\end{eqnarray}
for flavons in $\Phi^{\nu}$ sector. The VEVs of $\varphi$, $\eta$, $\chi$ and $\xi$ break $S_{4}$ completely, since acting on the vacua with
$T$ or $T^{2}$ the directions of them are invariant except an overall phase, while those of $\phi$ and $\Delta$ are invariant under the
four elements 1, $S^{2}$, $TST$ and $TSTS^{2}$. Moreover the order of magnitude for the VEVs $\langle\Phi^{e}\rangle$ and
$\langle\Phi^{\nu}\rangle$ are taken as $\lambda^{2}_{c}\Lambda$ and $\lambda^{3}_{c}\Lambda$, respectively. The reason for the constrains on
the order of the VEVs is that they should be responsible for the strong mass hierarchies of charged fermions. The small expansion parameters 
$\epsilon=\frac{\langle\Phi^{e}\rangle}{\Lambda}$ and $\delta=\frac{\langle\Phi^{\nu}\rangle}{\Lambda}$ are used in the following discussions.

\subsection{\textsl{Neutrino}}\label{S3sub1}
The right-handed neutrinos are SU(5) singlet \textbf{1} in the model, thus the light neutrino masses are only generated through type-I seesaw mechanism
\begin{equation}\label{eq:seesaw}
\qquad    m_{\nu}=-m^{T}_{D}M^{-1}_{M}m_{D}
\end{equation}
where the $m_{D}$ and $M_{M}$ are Dirac and Majorana mass matrices respectively. The two matrices are derived from the superpotential invariant under
the flavor symmetry. Concretely the superpotential in neutrino sector is as follows
\begin{equation}\label{eq:wneutrino}
\qquad  w_{\nu}=\frac{y_{\nu_{1}}}{\Lambda}(FN^{c})_{3_{1}}\phi H_{5}+\frac{y_{\nu_{2}}}{\Lambda}(FN^{c})_{2}\Delta H_{5}+\frac{1}{2}MN^{c}N^{c}
\end{equation}
The first two terms contribute to Dirac masses and the last one is Majorana righted-handed neutrinos mass. The Tri-Bimaximal mixing is reproduced
by the vacuum alignments of scalar fields  $\phi$ and $\Delta$, see Eq. (\ref{eq:Vf1}). Note that $\langle\phi\rangle$ and $\langle\Delta\rangle$
are invariant by acting the elements 1, $S^{2}$, $TST$ and $TSTS^{2}$ on them, hence the flavor symmetry is broken down to the Klein four subgroup.
After Electroweak and flavor symmetry breaking as Higgs field and flavons developing their VEVs, neutrinos will gain their masses.
For Dirac neutrino mass matrix at LO we have
\begin{equation}\label{eq:LO Dirac mass}
\qquad    m_{D}=\frac{y_{\nu_{1}}v_{\phi}v_{5}}{\Lambda}\left(
            \begin{array}{ccc}
              2 & -1 & -1 \\
              -1 & 2 & -1 \\
              -1 & -1 & 2 \\
            \end{array}
            \right)+\frac{y_{\nu_{2}}v_{\Delta}v_{5}}{\Lambda}\left(
            \begin{array}{ccc}
              0 & 1 & 1 \\
              1 & 1 & 0 \\
              1 & 0 & 1 \\
            \end{array}
            \right)
\end{equation}
and Majorana mass matrix is
\begin{equation}\label{eq:LO Majorana mass}
\qquad    M_{M}=\left(
        \begin{array}{ccc}
          M & 0 & 0 \\
          0 & 0 & M \\
          0 & M & 0 \\
        \end{array}
      \right)
\end{equation}
The eigenvalues of Majorana mass matrix $M_{M}$ can be diagonalized by unitary transformation
\begin{equation}\label{eq:UM}
\qquad    U^{T}_{R}M_{M}U_{R}=\textrm{diag}(M,M,M), \quad U_{R}=\left(
                     \begin{array}{ccc}
                       1 & 0 & 0 \\
                       0 & e^{i\alpha}/\sqrt{2} & -ie^{i\alpha}/\sqrt{2} \\
                       0 & e^{-i\alpha}/\sqrt{2} & ie^{-i\alpha}/\sqrt{2} \\
                     \end{array}
                   \right)
\end{equation}
where $\alpha$ is a phase parameter. All three right-handed neutrinos are degenerate with mass equal to M, the unitary transformation cannot be
solely determined.

Using Eq.~\eqref{eq:seesaw}, the derived light neutrino mass matrix can be diagonalized by the transformation
\begin{equation}\label{eq:diagonal neutrino mass}
\qquad    m^{\textrm{diag}}_{\nu}=U^{T}_{\nu}m_{\nu}U_{\nu}=\textrm{Diag}(m_{1},m_{2},m_{3})
\end{equation}
where the light neutrino masses $m_{1},m_{2},m_{3}$ are
\begin{eqnarray}\label{eq:mlight}
\qquad m_{1} = \Big|-\frac{(3a-b)^{2}}{M}\Big|\frac{v^{2}_{5}}{\Lambda^{2}},\quad
   m_{2} = \Big|-\frac{4b^{2}}{M}\Big|\frac{v^{2}_{5}}{\Lambda^{2}},\quad
   m_{3} = \Big|\frac{(3a+b)^{2}}{M}\Big|\frac{v^{2}_{5}}{\Lambda^{2}}
\end{eqnarray}
in which $a=y_{\nu_{1}}v_{\phi}$, $b=y_{\nu_{2}}v_{\Delta}$. While the unitary matrix $U_{\nu}$ in Eq.~\eqref{eq:diagonal neutrino mass} is given by
\begin{equation}\label{eq:Unu}
\qquad U_{\nu} = U_{TB}P_{\nu} = \left(
         \begin{array}{ccc}
          \sqrt{\frac{2}{3}} & \frac{1}{\sqrt{3}} & 0 \\
         -\frac{1}{\sqrt{6}} & \frac{1}{\sqrt{3}} & -\frac{1}{\sqrt{2}} \\
         -\frac{1}{\sqrt{6}} & \frac{1}{\sqrt{3}} & \frac{1}{\sqrt{2}} \\
         \end{array}
         \right)\left(
         \begin{array}{ccc}
           e^{i\vartheta_{1}/2} & 0 & 0 \\
           0 & e^{i\vartheta_{2}/2} & 0 \\
           0 & 0 & e^{i\vartheta_{3}/2} \\
         \end{array}
       \right)
\end{equation}
Thus the famous TBM mixing matrix is obtained at LO exactly. The phases $\vartheta_{i},i=1,2,3$ can be easily obtained as following
\begin{eqnarray}\label{eq:vartheta}
  \qquad \vartheta_{1}=\arg\left(-\frac{(3a-b)^{2}}{M}\frac{v^{2}_{5}}{\Lambda^{2}}\right)   ,
  \quad \vartheta_{2}=\arg\left(-\frac{4b^{2}}{M}\frac{v^{2}_{5}}{\Lambda^{2}}\right)    ,
  \quad \vartheta_{3}=\arg\left(\frac{(3a+b)^{2}}{M}\frac{v^{2}_{5}}{\Lambda^{2}}\right)
\end{eqnarray}
Note that the lepton PMNS mixing matrix is often parameterized by the standard PDG form as
\begin{equation}\label{eq:UPMNS}
 \qquad U_{\ell}=\left(
                   \begin{array}{ccc}
                     c_{12}c_{13} & s_{12}c_{13} & s_{13}e^{-i\delta_{13}} \\
                     -s_{12}c_{23}-c_{12}s_{23}s_{13}e^{i\delta_{13}} & c_{12}c_{23}-s_{12}s_{23}s_{13}e^{i\delta_{13}} & s_{23}c_{13} \\
                     s_{12}s_{23}-c_{12}c_{23}s_{13}e^{i\delta_{13}} & -c_{12}s_{23}-s_{12}c_{23}s_{13}e^{i\delta_{13}} & c_{23}c_{13} \\
                   \end{array}
                 \right)\left(
                 \begin{array}{ccc}
                 e^{-i\frac{\alpha_{1}}{2}} & 0 & 0 \\
                 0 & e^{-i\frac{\alpha_{2}}{2}} & 0 \\
                 0 & 0 & 1 \\
                 \end{array}
                 \right)
\end{equation}
where $c_{ij}=\cos\theta_{ij}$, $s_{ij}=\sin\theta_{ij}$ with $\theta_{ij}\in[0,\pi/2]$. The Dirac CP violating phase $\delta_{13}$, the two Majorana
CP violating phases $\alpha_{1}$ and $\alpha_{2}$ are all permitted to vary in the period of $0\sim2\pi$. In the following we shall identify the two
Majorana phases as $\alpha_{i}=\vartheta_{3}-\vartheta_{i}$ and express them in terms of the lightest neutrino mass.

Here let us leave the angles and phases for a moment and estimate the scale of \textsl{M} and the range of lightest neutrino mass in both hierarchies.
The mass square differences have been determined to high precision in several global neutrino data fits, such as~\cite{Garcia/2015fit}
\begin{equation}\label{eq:mass differences}
\qquad \Delta m^{2}_{sol} = 7.50^{+0.19}_{-0.197} \times 10^{-5}  \textrm{eV}^{2} , \quad
    \begin{cases}
    \Delta m^{2}_{atm} = 2.457^{+0.047}_{-0.047}  \times 10^{-3}  \textrm{eV}^{2}  \quad\;\; (\textrm{NH}) \\
    \Delta m^{2}_{atm} = -2.449^{+0.048}_{-0.047} \times 10^{-3}  \textrm{eV}^{2}  \quad (\textrm{IH}) \\
    \end{cases}
\end{equation}
where NH (IH) stands for normal (inverted) hierarchy of mass spectrum. Despite of the discrepancy of specific values and their accuracy,
another works of global fit data can be seen in Ref.~\cite{Forero/2012,Fogli/2012,Capozzi/2013,Forero/2014FIT}. The magnitude of masses $m_{i}$
can be roughly estimated around $10^{-2}$ $\textrm{eV}$ - $10^{-1}$ $\textrm{eV}$, and $v_{5}$ is electroweak sclale $\sim 10^{2}$ $\textrm{GeV}$,
then generally the scale of \textsl{M} will be
\begin{equation}\label{eq:scale M}
\qquad M \sim 10^{11\div 12} \, \textrm{GeV}.
\end{equation}

The parameters $a$ and $b$ are assumed to be of order $\lambda^{3}_{c}$, we can define the ratio as
\begin{equation}\label{eq:Rab}
 \qquad \frac{a}{b}=R e^{i\Theta}.
\end{equation}
then we can express the real parameters $R$ and $\Theta$ in terms of the light neutrino masses in Eq.~\eqref{eq:mlight} as
\begin{eqnarray}\label{eq:RTheta}
  \qquad R =\frac{1}{3} \sqrt{2\Big(\frac{m_{3}}{m_{2}}+\frac{m_{1}}{m_{2}}\Big)-1},
  \qquad \cos\Theta = \frac{\frac{m_{3}}{m_{2}}-\frac{m_{1}}{m_{2}}}{\sqrt{2\Big(\frac{m_{3}}{m_{2}}+\frac{m_{1}}{m_{2}}\Big)-1}}.
\end{eqnarray}
Taking into account the experimental values of two mass differences, cf. Eq.~\eqref{eq:mass differences}, we have only one free parameter left, which
can be chosen to be the lightest neutrino mass ($m_{1}$ in NH or $m_{3}$ in IH) for convenient. Imposing the constraint $|\cos\Theta|\leq1$, one could
obtain the limits for the lightest neutrino masses as
\begin{eqnarray}
  & & m_{1}\geq0.011 \textrm{eV}, \quad \textrm{NH}\nonumber\\
  & & m_{3}\geq0.028 \textrm{eV}, \quad \textrm{IH}
\end{eqnarray}
where only the best fit values are used in the estimation.

The Majorana phases in the PDG standard parameterization are naively defined as
\begin{equation}\label{eq:Majorana phases}
  \qquad \alpha_{1}=\vartheta_{3}-\vartheta_{1}, \quad \alpha_{2}=\vartheta_{3}-\vartheta_{2},
\end{equation}
inserting the expressions of $\vartheta_{i}$, $R$ and $\Theta$, cf. Eq.~\eqref{eq:vartheta} and Eq.~\eqref{eq:RTheta}, respectively, yield
\begin{eqnarray}
  & &  \sin\alpha_{1}=\frac{12R(1-9R^{2})\sin\Theta}{(1+9R^{2})^{2}-36R^{2}\cos^{2}\Theta},\qquad
  \cos\alpha_{1}=\frac{(1-9R^{2})^{2}-36R^{2}\sin^{2}\Theta}{(1+9R^{2})^{2}-36R^{2}\cos^{2}\Theta}\nonumber\\
  & &  \sin\alpha_{2}=\frac{6R\sin\Theta(1+3R\cos\Theta)}{1+9R^{2}+6R\cos\Theta},\qquad\quad
  \cos\alpha_{2}=\frac{1+9R^{2}\cos2\Theta+6R\cos\Theta}{1+9R^{2}+6R\cos\Theta}.
\end{eqnarray}
The two Majorana phases $\alpha_{1}$ and $\alpha_{2}$ can take two different sets of values in principle which corresponding to $\sin\Theta>0$ and
$\sin\Theta<0$, respectively. The reason is that the neutrino mass order which can be either NH or IH determines the sigh of $\cos\Theta$. The Dirac
CP phase $\delta_{13}$ is undetermined for the vanishing $\theta_{13}$ in TBM mixing.

Besides the operators in Eq.~\eqref{eq:wneutrino}, it is worth to consider effective operators, i.e., the higher dimensional Weinberg operators
\cite{Weinberg/1979} would also contribute to neutrino masses. In this model the effective operators for both Dirac and Majorana mass terms are
\footnote{The operators $FFH_{5}H_{5}$ and $FFH_{45}H_{45}$ represent $(F_{i})_{\alpha}(F_{j})_{\beta}H_{5}^{\alpha}H_{5}^{\beta}$
and $(F_{i})_{\alpha}(F_{j})_{\beta}(H_{45})^{\gamma\alpha}_{\delta}(H_{45})^{\delta\beta}_{\gamma}$ respectively, where the Greek indices denote the
SU(5) tensor contractions, the Latin indices are the fermion generations or say contractions in $S_{4}$ space.}
\begin{eqnarray}\label{eq:weffective}
& &    W^{eff}_{\nu}=\frac{y_{1}}{\Lambda_{W}}FFH_{5}H_{5}
       +\sum_{i=1}^{3}\frac{y^{(1_{1})}_{i}}{\Lambda^{3}_{W}}FFH_{5}H_{5}\mathcal{O}^{1_{1}}_{i}
       +\sum_{i=1}^{2}\frac{y^{(2)}_{i}}{\Lambda^{3}_{W}}FFH_{5}H_{5}\mathcal{O}^{(2)}_{i}\nonumber\\
& &    \qquad\qquad+\sum_{i=1}^{2}\frac{y^{(3_{1})}_{i}}{\Lambda^{3}_{W}}FFH_{5}H_{5}\mathcal{O}^{(3_{1})}_{i}
       +\frac{y_{2}}{\Lambda^{4}_{W}}FFH_{45}H_{45}(\phi\Delta)_{3_{2}}\zeta,
\end{eqnarray}
where the operators $\mathcal{O}_{i}$ represent the following specific $S_{4}$ contractions of flavons in $\Phi^{\nu}$
\begin{eqnarray}
  \qquad \mathcal{O}^{1_{1}}_{i}=\{(\phi\phi)_{1_{1}},(\Delta\Delta)_{1_{1}},\zeta\zeta\},
  \qquad \mathcal{O}^{2}_{i}=\{(\phi\phi)_{2},(\Delta\Delta)_{2}\},
  \qquad \mathcal{O}^{3_{1}}_{i}=\{(\phi\phi)_{3_{1}},(\phi\Delta)_{3_{1}}\}
\end{eqnarray}

With the VEVs of Higgs fields, the operators generate the neutrino mass terms, in unit of $\frac{v^{2}_{5}}{\Lambda_{W}}$ as follows
\begin{equation}\label{eq:mW}
\qquad    m_{W}=y^{(1_{1})}
    \left(
    \begin{array}{ccc}
    1 & 0 & 0 \\
    0 & 0 & 1 \\
    0 & 1 & 0 \\
    \end{array}
    \right)
    +y^{(2)}\left(
    \begin{array}{ccc}
    0 & 1 & 1 \\
    1 & 1 & 0 \\
    1 & 0 & 1 \\
    \end{array}
    \right)
    +y^{(3_{1})}\left(
    \begin{array}{ccc}
    2 & -1 & -1 \\
    -1 & 2 & -1 \\
    -1 & -1 & 2 \\
    \end{array}
    \right)
\end{equation}
in which the coefficients are
\begin{eqnarray}
 \qquad y^{(1_{1})}=y_{1}+3y^{(1_{1})}_{1}\frac{v^{2}_{\phi}}{\Lambda^{2}_{W}}+2y^{(1_{1})}_{2}\frac{v^{2}_{\Delta}}{\Lambda^{2}_{W}}
      +y^{(1_{1})}_{3}\frac{v^{2}_{\zeta}}{\Lambda^{2}_{W}},
  \quad y^{(2)}=3y^{(2)}_{1}\frac{v^{2}_{\phi}}{\Lambda^{2}_{W}}+\frac{v^{2}_{\Delta}}{\Lambda^{2}_{W}},
  \quad y^{(3_{1})}=2y^{(3_{1})}_{2}\frac{v_{\phi}v_{\Delta}}{\Lambda^{2}_{W}}
\end{eqnarray}

One can realize the structure of $m_{W}$ is exactly the same as that of $m_{D}$ in Eq.~\eqref{eq:LO Dirac mass} and of $M_{M}$ in
Eq.~\eqref{eq:LO Majorana mass} combined, hence it can be exactly diagonalized by TBM mixing matrix
\begin{equation}\label{eq:effective mass}
\qquad    m^{\textrm{diag}}_{W}=U^{T}_{TB}m_{W}U_{TB}=\textrm{Diag}(m_{W1},m_{W2},m_{W3}),
\end{equation}
where the light effective neutrino masses $m_{W1}$, $m_{W2}$ and $m_{W3}$ come from the Weinberg operators, they are given by
\begin{eqnarray}
  \qquad m_{W1}= y^{(1_{1})}-y^{(2)}+3y^{(3_{1})},
  \quad m_{W2}= y^{(1_{1})}+2y^{(2)},
  \quad m_{W3}= -y^{(1_{1})}+y^{(2)}+3y^{(3_{1})}
\end{eqnarray}

Obviously we can compare the
relative magnitude between $m^{\textrm{diag}}_{\nu}$ and $m^{\textrm{diag}}_{W}$ with the assumption that all couplings, $y_{\nu_{1,2}}$,
and $y^{c}$, are of order one, then the  ratio could be
\begin{equation}\label{eq:ratio}
\qquad    \frac{m_{Wi}}{m_{i}}\sim \frac{M\Lambda^{2}}{\Lambda_{W}v^{2}_{\phi}}\sim 10^{4}\frac{M}{\Lambda_{W}}\sim \frac{\Lambda_{GUT}}{\Lambda_{W}}
\end{equation}
The contribution to neutrino mass would be larger than the seesaw one if the Weinberg operator has cutoff $\Lambda_{W}\sim \Lambda_{GUT}$. In order
to avoid the problem we will require $\Lambda_{W}\gg\Lambda_{GUT}$, such that $\Lambda_{W}\sim\Lambda_{Planck}$, then the 5-dimensional effective
operator can be neglected.
\subsection{\textsl{Up-type quarks}}\label{S3sub2}
The masses of up-type quarks are generated by $S_{4}$ symmetry breaking in the invariant superpotential at LO. The enormous mass hierarchies
among up-type quarks should be guaranteed by the VEVs of scalar flavons. At LO the invariant superpotential under the whole symmetry groups
is simply as \footnote{\; The product $T_{i}T_{j}H_{5}$
denotes $\varepsilon_{\alpha\beta\gamma\rho\sigma}T^{\alpha\beta}_{i}T^{\gamma\rho}_{j}H^{\sigma}_{5}$, and $T_{i}T_{j}H_{45}$
denotes $\varepsilon_{\alpha\beta\gamma\rho\sigma}T^{\alpha\beta}_{i}T^{\zeta\gamma}_{j}(H_{45})^{\rho\sigma}_{\zeta}$, where the Greek
indices indicates the SU(5) tensor contractions and, Latin indices $i,j =1,2,3$, are family indices. Here we don't write the $S_{4}$
contractions because all contractions in $S_{4}$ space should be first reduced to that in SU(5) space. The SU(5) tensor contractions are more
fundamental ones for the calculations, the Clebsch-Gordan coefficients in Eq.~\eqref{eq:LO up matrix} are mainly determined by those SU(5) tensor
contractions.}
\begin{eqnarray}\label{eq:wUP}
& & w_{U} = y_{t}T_{3}T_{3}H_{5}+\frac{y_{c}}{\Lambda^{2}}T_{2}T_{2}\sigma\vartheta H_{5}
    +\frac{y_{ct}}{\Lambda}T_{2}T_{3}\sigma H_{45}+\frac{y_{u}}{\Lambda^{3}}T_{1}T_{1}H_{5}\eta\Delta\zeta \nonumber\\
& & \qquad +\sum_{i=1}^{3} \frac{y'_{ti}}{\Lambda^{2}}T_{3}T_{3}H_{5}\mathcal{O}^{U1}_{i}
    +\sum_{i=1}^{3} \frac{y'_{cti}}{\Lambda^{3}}T_{2}T_{3}H_{45}\mathcal{O}^{U2}_{i}
\end{eqnarray}
where the operators read
\begin{equation}\label{eq:ou1}
  \qquad \mathcal{O}^{U1}_{i}=\{\zeta\zeta, \Delta\Delta, \phi\phi\},
  \quad \mathcal{O}^{U2}_{i}=\{\sigma\zeta\zeta, \sigma\Delta\Delta, \sigma\phi\phi\}\\
\end{equation}
Then after the scalar fields develop their VEVs in Eqs.~\eqref{eq:Ve11} and~\eqref{eq:Ve12} and gauge symmetry breaking, the mass matrix 
of up-type quarks can be written as
\begin{equation}\label{eq:LO up matrix}
\qquad M_{U}=\left(
  \begin{array}{ccc}
  -8y_{u}\frac{v_{\eta}v_{\Delta}v_{\zeta}}{\Lambda^{3}}v_{5} & 0 & 0\\
  0 & 8y_{c}\frac{v_{\sigma}v_{\vartheta}}{\Lambda^{2}}v_{5}
  & 8y_{ct}\frac{v_{\sigma}}{\Lambda}v_{45} \\
  0
  & -8y_{ct}\frac{v_{\sigma}}{\Lambda}v_{45} & 8y_{t}v_{5}\\
  \end{array}
  \right)+8\left(
            \begin{array}{ccc}
              0 & 0 & 0 \\
              0 & 0 & y'_{ct}\epsilon\delta^{2}v_{45} \\
              0 & -y'_{ct}\epsilon\delta^{2}v_{45} & y'_{t}\delta^{2}v_{5} \\
            \end{array}
          \right)
\end{equation}
Then mass matrix can be diagonalized by the bi-unitary transformation
\begin{equation}\label{eq:Dmu}
\qquad m_{U}=U^{\dag}_{R}M_{U}U_{L}=\textrm{Diag}(m_{u},m_{c},m_{t})
\end{equation}
in which the mass eigenvalues are
\begin{eqnarray}\label{eq:up masses}
  \qquad m_{u} = \Big|-8y_{u}\frac{v_{\eta}v_{\Delta}v_{\zeta}}{\Lambda^{3}}v_{5}\Big|,
  \qquad m_{c} = \Big|8y_{c}\frac{v_{\sigma}v_{\vartheta}}{\Lambda^{2}}v_{5}
     +8\frac{y^2_{ct}}{y_{t}}\frac{v^{2}_{\sigma}}{\Lambda^{2}}\frac{v^2_{45}}{v_{5}}\Big| ,
  \qquad m_{t} = |8y_{t}v_{5}|
\end{eqnarray}
where the small contributions from the second matrix in Eq.~\eqref{eq:LO up matrix} is safely dropped or say reabsorbed into the redefinition
of the couplings $y_{t}$ and $y_{ct}$. In fact all the correctional contributions to the (33) element can be reabsorbed into the $y_{t}$ whose
order of magnitude is not changed. We can summarize that all the masses of up-type quarks are obtained at LO, particularly the top quark mass
is produced at tree level. The mass hierarchy between charm and top quark is obtained given that the $v_{\sigma}$ and $v_{\vartheta}$ of
order $\lambda^{2}_{c}\Lambda$. The up quark is also produced the correct order by means of $v_{\eta}\sim\lambda^{2}_{c}\Lambda$ and
$v_{\Delta, \zeta}\sim \lambda^{3}_{c}\Lambda$. The mass hierarchies are obtained as usual: $m_{u}:m_{c}:m_{t}=\lambda^{8}_{c}:\lambda^{4}_{c}:1$.
Due to the quantity $\lambda^{8}_{c}$ is only comprised of the combination $\epsilon\delta^{2}$ rather than $\epsilon^{4}$, we can conclude that
the hierarchical VEVs of $\Phi^{e}$ and $\Phi^{\nu}$ are essential in the model, as declared in section~\ref{S2}.

After diagonalize the mass matrix~\eqref{eq:LO up matrix}, one can note that the mixing of up-type quarks only exists between charm quark and
top quark, which is an experimental acceptable feature of our model. Thus the form of the resulting mixing matrix $U_{L}$ is rather simple
\begin{equation}\label{eq:UL}
  \qquad U_{L}=\left(
          \begin{array}{ccc}
            1 & 0 & 0 \\
            0 & 1 & S^{u}_{23} \\
            0 & -S^{u\ast}_{23} & 1 \\
          \end{array}
        \right)
\end{equation}
with the mixing angles
\begin{eqnarray}
  \qquad S^{u}_{12} = S^{u}_{13} = 0,
  \qquad S^{u}_{23} = -(\frac{y_{ct}}{y_{t}}\frac{v_{\sigma}}{\Lambda}\frac{v_{45}}{v_{5}})^{\ast},
\end{eqnarray}
where $S^{u}_{ij}$ is the sine of mixing angles and $C^{u}_{ij} \simeq 1$ is assumed. 
In the following chapter we may aware of the (23) and (32) elements of CKM mixing matrix are totally determined by $S^{u}_{23}$.
\subsection{\textsl{Down-type quarks and charged leptons}}\label{S3sub3}
The matter fields appeared in the superpotential $w_{U}$ in Eq.~\eqref{eq:wUP} which gives rise to the masses of up-type quarks only include
the SU(5) \textbf{10}-dimension representational $T_{i}$ ($i=1,2,3$), while down-type quarks and charged leptons are not the case.
The down-type quark fields (and charged lepton fields) and their conjugate fields are assigned to be \textbf{10} and $\bar{\textbf{5}}$
representations of SU(5), respectively. Consequently the superpotential of down-type quarks and charged leptons is comprised of
$T_{i}$ and $F$ together with the down-type Higgs fields $H_{\bar{5},\overline{45}}$ and the flavons. In addition the adjoint field $H_{24}$
plays a important role to gain the novel CG factors that can lead to a desired GUT relation between masses and mixing angle, as elucidated
in section~\ref{S2}. To be specific the LO superpotential which gives rise to the masses of down-type quarks and charged leptons
is~\footnote{\; Similarly it is easy to write the basic contractions of these operators in SU(5) space:
$T_{i}F_{j}H_{\bar{5}}=T_{i}^{\alpha\beta}(F_{j})_{\alpha}(H_{\bar{5}})_{\beta}$ and,
$T_{i}F_{j}H_{\overline{45}}=T_{i}^{\alpha\beta}(F_{j})_{\gamma}(H_{\overline{45}})^{\gamma}_{\alpha\beta}$. The index structures of the
contractions can be found, e.g., in~\cite{Duque/2008Fmass}.}
\begin{eqnarray}\label{eq:wdown}
& &  w_{D} = \frac{y_{b}}{\Lambda}T_{3}F\varphi H_{\bar{5}}
    +\sum_{i=1}^{2}\frac{y_{si}}{\Lambda^{2}}T_{2}F\mathcal {O}^{D1}_{i}H_{\overline{45}}
    + \frac{y_{bd}}{\Lambda^{2}}T_{1}F\varphi\sigma H_{\overline{45}}
    +\frac{y'_{s}}{\Lambda^{3}}T_{2}F(\varphi\eta)_{3_{2}}\sigma H_{\overline{45}}   \nonumber \\
& & \qquad+\sum_{i=1}^{2}\frac{f_{di}}{\langle H_{24}\rangle^{3}}T_{1}F\mathcal{O}^{D2}_{i} H_{\bar{5}}
     +\sum_{i=1}^{3}\frac{g_{di}}{\Lambda^{2}\langle H_{24}\rangle}T_{2}F\mathcal{O}^{D3}_{i} H_{\bar{5}}
    +\sum_{i=1}^{6}\frac{h_{di}}{\Lambda^{3}}T_{3}F\mathcal {O}^{D4}_{i} H_{\bar{5}}+\cdots
\end{eqnarray}
where dots stand for higher order operators, and the operators $\mathcal {O}^{D}$ are
\begin{eqnarray}
  & & \mathcal{O}^{D1}_{i} = \{ \chi\sigma, \xi\zeta \},   \quad \mathcal{O}^{D2}_{i} = \{\chi\xi\rho, \chi\xi\xi \},\nonumber \\
  & & \mathcal{O}^{D3}_{i} =  \{ \phi\phi\phi, \phi\phi\Delta, \phi\Delta\Delta\}, \quad
 \mathcal{O}^{D4}_{i} = \{\varphi\phi\phi, \varphi\phi\Delta, \varphi\Delta\Delta, \varphi\zeta\zeta, \eta\phi\phi, \eta\Delta\phi \}.
\end{eqnarray}
The operators involved $\mathcal{O}^{D}$ in Eq.~\eqref{eq:wdown} should include all possible independent $S_{4}$ contractions. The down-type quarks and
charged leptons in the superpotential Eq.~\eqref{eq:wdown} would obtain their masses as the Higgs fields and flavon fields developing their VEVs after
the symmetries are broken. Note that the operators involved $\mathcal{O}^{D3}_{i}$ have vanishing contribution to the entries of Yukawa matrix.
With the VEVs of the flavons and Higgses, the mass matrix of down-type quarks is immediately derived as follows
\begin{equation}\label{eq:down mass matrix}
    \qquad M_{D}=\left(
    \begin{array}{ccc}
    0
    & y^{d}_{12}\epsilon\delta v_{\overline{45}}
    &y^{d}_{13}\epsilon \delta^{2}v_{\bar{5}}
    \\
    y^{d}_{21}\epsilon^{3} v_{\bar{5}}
    & (y^{d}_{22}+y^{d'}_{22}\epsilon) \epsilon^{2} v_{\overline{45}}
    &y^{d}_{23}\epsilon\delta^{2}v_{\bar{5}}
    \\
    y^{d}_{31}\epsilon^{2} v_{\overline{45}}
    & 0
    & y^{d}_{33}\epsilon v_{\bar{5}}+y^{d'}_{33}\epsilon\delta^{2}v_{\bar{5}}
    \\
    \end{array}
    \right)
\end{equation}
where the coefficients $y^{d}_{ij}$ ($i, j$=1, 2, 3) and those with primes are linear combinations of LO coefficients.
The (21) element implies $v_{\Phi^{e}}/\langle H_{24}\rangle\sim\epsilon$, or equivalently $\langle H_{24}\rangle\sim\Lambda$
\footnote{\; The adjoint field $H_{24}$ has the VEV along the direction
$\langle H_{24}\rangle=\sqrt{\frac{2}{15}}v_{24} \textrm{diag}(1,1,1,-\frac{3}{2},-\frac{3}{2})$, which is responsible for the broken of SU(5) GUT
symmetry down to the standard model symmetry $SU(3)_{c}\times SU(2)_{L}\times U(1)_{Y}$.}
which is an essential condition to obtain the proper order of magnitude for the masses of down quark and electron, and hence for the hierarchies
of all the masses. Similarly the mass matrix of charged leptons which is equivalent to the transposed $M_{D}$ is given as
\begin{equation}\label{eq:lepton mass matrix}
    \qquad M_{\ell}=\left(
    \begin{array}{ccc}
     0
     & -\frac{8}{27}y^{d}_{21}\epsilon^{3} v_{\bar{5}}
     & -3y^{d}_{31}\epsilon^{2} v_{\overline{45}} \\
     -3y^{d}_{12}\epsilon\delta v_{\overline{45}}
     & -3(y^{d}_{22}+y^{d'}_{22}\epsilon)\epsilon^{2} v_{\overline{45}}
     & 0\\
       y^{d}_{13}\epsilon \delta^{2}v_{\bar{5}}
     & y^{d}_{23}\epsilon\delta^{2}v_{\bar{5}}
     & y^{d}_{33}\epsilon v_{\bar{5}}+y^{d'}_{33}\epsilon\delta^{2}v_{\bar{5}}
     \\
    \end{array}
    \right)
\end{equation}
The CG coefficients which manifest in the entries of the mass matrices are determined by the way of the tensors contracted in SU(5) space.
The two mass matrices $M_{D}$ and $M_{\ell}$ can be diagonalized by the similar bi-unitary transformations as in up quark sector
\begin{equation}\label{eq:Udl}
    \qquad V^{D\dag}_{R}M_{D}D_{L}=\textrm{diag}(m_{d},m_{s},m_{b}), \quad V^{\ell\dag}_{R}M_{\ell}V^{\ell}_{L}=\textrm{diag}(m_{e},m_{\mu},m_{\tau}).
\end{equation}
The mass eigenvalues of down type quarks are
\begin{eqnarray}\label{eq:down masses}
  \qquad m_{d} \simeq \Big|-\frac{y^{d}_{21}y^{d}_{12}}{y^{d}_{22}}\epsilon^{2}\delta v_{\bar{5}}\Big|,
  \quad m_{s} \simeq \Big|(y^{d}_{22}+y^{d'}_{22}\epsilon)\epsilon^{2}v_{\overline{45}}
  +\frac{y^{d}_{21}y^{d}_{12}}{y^{d}_{22}}\epsilon^{2}\delta v_{\bar{5}}\Big|  ,
  \quad m_{b} \simeq |y^{d}_{33}\epsilon v_{\overline{45}}|
\end{eqnarray}
and those of charged leptons are
\begin{eqnarray}\label{eq:lepton masses}
  \qquad m_{e} \simeq \Big|\frac{8}{27}\frac{y^{d}_{21}y^{d}_{12}}{y^{d}_{22}}\epsilon^{2}\delta v_{\bar{5}}\Big|,
  \quad m_{\mu} \simeq \Big|-3(y^{d}_{22}+y^{d'}_{22}\epsilon)\epsilon^{2}v_{\overline{45}}
  -\frac{8y^{d}_{21}y^{d}_{12}}{27y^{d}_{22}}\epsilon^{2}\delta v_{\bar{5}}\Big|,
  \quad m_{\tau} \simeq |y^{d}_{33}\epsilon v_{\bar{5}}|
\end{eqnarray}
From the mass expressions in Eq.~\eqref{eq:down masses} and Eq.~\eqref{eq:lepton masses}, one can easily find that bottom quark and tau
lepton have the same mass, and the mass of muon is three times of that of strange quark, and the mass of
electron is $\frac{8}{27}$ of that of down quark
\begin{equation}\label{eq:downmass}
 \qquad m_{\tau} \simeq m_{b}, \qquad m_{\mu} \simeq 3m_{s}, \qquad m_{e} \simeq \frac{8}{27}m_{d},
\end{equation}
The bottom-tau unification and Georgi-Jarlskog relation~\cite{GJ/1979} are produced in the model, and a new novel mass ratio $\frac{m_{e}}{m_{d}}$
which is favored in phenomenology is obtained. The above mass relations between down-type quarks and charged leptons give rise to the combined
relation $\frac{m_{\mu}}{m_{s}}\frac{m_{d}}{m_{e}}\simeq 10.1$, which is well within the $1\sigma$ range discussed in~\cite{Antusch/2013RUN},
i.e., the double ratio of Yukawa couplings at the scale $M_{GUT}$
\begin{equation}\label{eq:dbratio}
\qquad \frac{y_{\mu}}{y_{s}}\frac{y_{d}}{y_{e}}\approx 10.7\pm^{1.8}_{0.8}.
\end{equation}
In contrast, the double ratio in the distinguished original George-Jarlskog relation~\cite{GJ/1979}, which implies
$\frac{y_{\mu}}{y_{s}}\frac{y_{d}}{y_{e}}=9$, deviates from the phenomenological favored result more than $2\sigma$.

The unitary transformation matrices $D_{L}$ and $V^{\ell}_{L}$ are approximately as
\begin{equation}\label{eq:DL}
    \qquad D_{L}=\left(
    \begin{array}{ccc}
    1 & (\dfrac{y^{d}_{21}}{y^{d}_{22}}\dfrac{v_{\bar{5}}}{v_{\overline{45}}}\epsilon)^{\ast} &
    (\dfrac{y^{d}_{31}}{y^{d}_{33}}\dfrac{v_{\overline{45}}}{v_{\bar{5}}}\epsilon)^{\ast} \\
    -\dfrac{y^{d}_{21}}{y^{d}_{22}}\dfrac{v_{\bar{5}}}{v_{\overline{45}}}\epsilon & 1 &
    0\\
    -\dfrac{y^{d}_{31}}{y^{d}_{33}}\dfrac{v_{\overline{45}}}{v_{\bar{5}}}\epsilon &
    (\dfrac{y^{d\ast}_{21}}{y^{d\ast}_{22}}\dfrac{y^{d}_{31}}{y^{d}_{33}})|\epsilon|^{2}
    & 1 \\
    \end{array}
    \right)
\end{equation}
\begin{equation}\label{eq:VL}
   \qquad V^{\ell}_{L}=\left(
   \begin{array}{ccc}
   1 & (\dfrac{y^{d}_{12}}{y^{d}_{22}}\dfrac{\delta}{\epsilon})^{\ast} & 0\\
   -\dfrac{y^{d}_{12}}{y^{d}_{22}}\dfrac{\delta}{\epsilon} & 1 & 0 \\
   0 & 0 & 1 \\
   \end{array}
   \right)
\end{equation}
The complete quark mixing matrix $V_{CKM}$ is composed of mixing matrices of both up-type quark and down-type quark sector
\begin{equation}\label{eq:CKM}
    \qquad V_{CKM}=U^{\dag}_{L}D_{L}
\end{equation}
then we can directly get all elements of CKM matrix
\begin{eqnarray}\label{eq:CKMelements}
 & & V_{ud} \simeq V_{cs} \simeq V_{tb} \simeq 1 ,
 \qquad V^{\ast}_{us} \simeq -V_{cd} \simeq \frac{y^{d}_{21}}{y^{d}_{22}}\dfrac{v_{\bar{5}}}{v_{\overline{45}}}\epsilon,
 \qquad V^{\ast}_{ub} \simeq \frac{y^{d}_{31}}{y^{d}_{33}}\frac{v_{\overline{45}}}{v_{\bar{5}}}\epsilon                                     \nonumber\\
 & & V_{td} \simeq -\frac{y^{d}_{31}}{y^{d}_{33}}\frac{v_{\overline{45}}}{v_{\bar{5}}}\epsilon
                      -\frac{y^{d}_{21}}{y^{d}_{22}}\dfrac{v_{\bar{5}}}{v_{\overline{45}}}
                      \frac{y_{ct}}{y_{t}}\frac{v_{\sigma}}{\Lambda}\frac{v_{45}}{v_{5}}\epsilon,
 \qquad V^{\ast}_{cb} \simeq -V_{ts} \simeq
 \frac{y_{ct}}{y_{t}}\frac{v_{\sigma}}{\Lambda}\frac{v_{45}}{v_{5}}
\end{eqnarray}
It is an experimental constrains that $V_{us}$ and $V_{cd}$ are Cabibbo angle $\lambda_{c}$, $V_{ub}$ and $V_{td}$ are of order $\lambda^{3}_{c}$,
which all demands a fine tuning between $v_{\overline{45}}$ and $v_{\bar{5}}$: $v_{\overline{45}}\sim \lambda_{c}v_{\bar{5}}$. Adding this condition,
we can easily check that the quark CKM mixing matrix is produced correctly. The Cabibbo angle are determined by the mixing between the first and
second family down-type quarks, the parameters $y^{d}_{21}$ and $y^{d}_{22}$ are of order one was assumed. $V_{cb}$ and $V_{ts}$ are determined
by mixing between second and third generation left-handed up quarks .

Similarly the resulting lepton mixing matrix $U_{PMNS}$ is written as
\begin{equation}\label{eq:PMNS}
    \qquad U_{PMNS}=V^{\ell\dag}_{L}U_{\nu}, \quad U_{\nu}=U_{TB}P_{\nu}
\end{equation}
The desired CKM-like mixing matrix $V^{\ell}_{L}$ in Eq.~\eqref{eq:VL} implies a large mixing angle between the first and the second generation of
charged leptons, and it will remarkably change the lepton mixing, although the TBM mixing is exactly produced in neutrino sector. After simple straight calculation, we arrive at the three leptonic mixing angles $\theta^{PMNS}_{ij}$ at LO as following
\begin{eqnarray}\label{eq:LO angles}
  & & \sin^{2} \theta^{PMNS}_{12} = \frac{|(U_{PMNS})_{e2}|^{2}}{1-|(U_{PMNS})_{e3}|^{2}} \nonumber\\
  & & \qquad\qquad\quad=\frac{1}{3}-\frac{2}{3}Re\Big(\frac{y^{d}_{12}}{y^{d}_{22}}\frac{\delta}{\epsilon}\Big)
  +\frac{1}{2}\Big|\frac{y^{d}_{12}}{y^{d}_{22}}\frac{\delta}{\epsilon}\Big|^{2}
  -\frac{1}{3}\Big|\frac{y^{d}_{12}}{y^{d}_{22}}\frac{\delta}{\epsilon}\Big|^{2}Re\Big(\frac{y^{d}_{12}}{y^{d}_{22}}\frac{\delta}{\epsilon}\Big)
\nonumber\\
  & & \sin^{2} \theta^{PMNS}_{23} = \frac{|(U_{PMNS})_{\mu3}|^{2}}{1-|(U_{PMNS})_{e3}|^{2}} =
  \frac{1}{2}\left(1+\frac{1}{2}\Big|\frac{y^{d}_{12}}{y^{d}_{22}}\frac{\delta}{\epsilon}\Big|^{2}\right)                              \nonumber\\
  & & \sin \theta^{PMNS}_{13} = |(U_{PMNS})_{e3}| =
  \frac{1}{\sqrt{2}}\Big|\frac{y^{d}_{12}}{y^{d}_{22}}\frac{\delta}{\epsilon}\Big|
\end{eqnarray}
As elucidated in Section~\ref{S2}, the ratio of CG coefficients $|\frac{c_{c}}{c_{b}}|\sim\mathcal{O}(2/\lambda_{c})$ is required to obtain
the large angle $\theta^{e}_{12}\sim\lambda_{c}$ in the GUT relation with the phenomenological mass ratio $\frac{m_{e}}{m_{\mu}}$
and $\theta^{d}_{12}$. In the present model the value of the ratio $|\frac{c_{c}}{c_{b}}|$ is $\frac{81}{8}$, which satisfies the requirement of order
$\mathcal{O}(2/\lambda_{c})$. Then we can conclude that the mixing matrix $V^{\ell}_{L}$ is accurate enough to hold
$\theta^{e}_{12}\sim\lambda_{c}$ and hence the empirical relation $\theta^{PMNS}_{13}\simeq\lambda_{c}/\sqrt{2}$ is obtained.

The large $\theta^{PMNS}_{13}$ arises from the contribution of charged lepton sector, and $\theta^{PMNS}_{12}$ prominent deviates from its TBM value
in the case. Since the reactor experiments have showed that lepton mixing angle $\theta^{PMNS}_{13}$ are about $\lambda_{c}/\sqrt{2}$, which
meant $\ds|\frac{y^{d}_{12}}{y^{d}_{22}}\frac{\delta}{\epsilon}| \sim \lambda_{c}$, then the deviations of three leptonic mixing angles
from their TBM values are roughly estimated as follows
\begin{eqnarray}\label{eq:sum rules}
\qquad \sin \theta^{PMNS}_{13} \sim \frac{\lambda_{c}}{\sqrt{2}} ,
\qquad \Big|\sin^{2} \theta^{PMNS}_{12}-\frac{1}{3}\Big| \sim \frac{2}{3}\lambda_{c},
\qquad \Big|\sin^{2} \theta^{PMNS}_{23}-\frac{1}{2}\Big| \sim \frac{\lambda^{2}_{c}}{4}
\end{eqnarray}
The relations are compatible with leptonic mixing sum rules~\cite{Plent/2005}. Recall that the experimental value
of $\theta^{PMNS}_{12}$~\cite{Garcia/2015fit} is very close to TBM value, the large departure is seemingly unsuitable. Setting the expansion
parameters $\epsilon$ and $\delta$ to be positive real numbers for simplicity, the problem could be settled by taking into account of a Dirac CP
violating phase, which is in fact the complex phase, denoted by $\phi_{12}$, of the ratio $y^{d}_{12}/y^{d}_{22}$. We remind the readers
that $\phi_{12}$ is in fact determined by $\arg(y^{d}_{12}\delta/y^{d}_{22}\epsilon$), since the expansion parameters
$\delta=\langle\Phi^{\nu}\rangle/\Lambda$ and $\epsilon=\langle\Phi^{e}\rangle/\Lambda$ are in general complex numbers as $y^{d}_{ij}$s.
The phases of  $\delta$ and $\epsilon$, however, can be incorporated by the phases of $y^{d}_{ij}$s. Ignoring the higher order deviations,
the deviation of angle $\theta^{PMNS}_{12}$ from the TBM prediction is approximately expressed as following
\begin{equation}\label{eq:12phi}
\qquad \sin^{2} \theta^{PMNS}_{12}-\frac{1}{3} \sim -\frac{2\sqrt{2}}{3}\Big|(U_{PMNS})_{e3}\Big|\cos \phi_{12},
\qquad \phi_{12}=\arg(\frac{y^{d}_{12}}{y^{d}_{22}})
\end{equation}
The correlation has been given in~\cite{Plent/2005} as mentioned before, and similar result with minus sign difference (because of different
diagonalization conventions of fermion mass matrix) has been obtained in Ref.~\cite{Marzo/2011}. In order to be consistent with the TBM
value, the phase $\phi_{12}$ should be around $\pi/2$ or $-\pi/2$ so that the sizable departure vanishes, or at least decreases to be of order
$\lambda^{2}_{c}$. The detailed study about the link between CP violation and charged lepton corrections to mixing angles is beyond the scope of the
present work, one may refer~\cite{Antusch/2005} for example.

\section{Vacuum alignment}\label{S4}
The vacuum alignment would be discussed as the natural solution of the scalar potential. The problem can be solved by so-called supersymmetric
driving field method introduced by Altarelli and Feruglio in Ref.~\cite{Alta/2006}. At LO, the superpotential of driving fields, which is
invariant under the flavor symmetry $S_{4}\times Z_{4} \times Z_{6} \times Z_{5}\times Z_{2}$, is given by
$w_{d}=w^{e}_{d}(\varphi_{0},\chi_{0},\eta_{0},\sigma_{0}, \xi_{0})+w^{\nu}_{d}(\phi_{0},\Delta_{0},\zeta_{0})$ where
\begin{eqnarray}\label{eq:LO wd}
    & & w_{d} = g_{1}\varphi_{0}\varphi\varphi+g_{2}(\varphi_{0}\varphi)_{2}\eta
    +h_{1}\eta_{0}(\varphi\varphi)_{1_{1}}+h_{2}\eta_{0}(\eta\eta)_{1_{1}}
    +g_{3}M_{\chi}\chi_{0}\chi+g_{4}\chi_{0}(\varphi\eta)_{3_{2}} \nonumber\\
    & &\qquad 
    +q_{1}\sigma_{0}\sigma\sigma+q_{2}\sigma_{0}\vartheta\vartheta
    +r_{1}\xi_{0}\xi\xi+r_{2}\xi_{0}\xi\rho \nonumber\\
    & &\qquad +f_{1}\phi_{0}(\phi\Delta)_{3_{2}}+f_{2}\Delta_{0}(\phi\phi)_{2}+f_{3}\Delta_{0}\Delta\Delta
    +f_{4}\zeta_{0}\zeta\zeta+f_{5}\zeta_{0}(\phi\phi)_{1_{1}}+f_{6}\zeta_{0}(\Delta\Delta)_{1_{1}}.
\end{eqnarray}
The vacuum alignments of all flavons in $\Phi^{e}$ and $\Phi^{\nu}$ are determined by deriving $w_{d}$ with respect to each component of the
driving fields $\Phi_{0}$ in SUSY limit. After minimized the derivative equations and solved each unknown component of all flavons,
the VEVs structures of the flavons can be obtained. Usually the solutions are not uniquely determined, we should choose one set by taking
into account some constrained conditions. The detailed minimization equations of flavons $\varphi, \eta$ and $\chi$
\begin{eqnarray}\label{eq:wde11}
  & & \frac{\partial w_{d} }{\partial \varphi_{01}} =
  2g_{1}(\varphi^{2}_{1}-\varphi_{2}\varphi_{3})+g_{2}(\eta_{1}\varphi_{2}+\eta_{2}\varphi_{3})=0\nonumber\\
  & & \frac{\partial w_{d} }{\partial \varphi_{02}} =
  2g_{1}(\varphi^{2}_{2}-\varphi_{1}\varphi_{3})+g_{2}(\eta_{1}\varphi_{1}+\eta_{2}\varphi_{2})=0\nonumber\\
  & & \frac{\partial w_{d} }{\partial \varphi_{03}} =
  2g_{1}(\varphi^{2}_{3}-\varphi_{1}\varphi_{2})+g_{2}(\eta_{1}\varphi_{3}+\eta_{2}\varphi_{1})=0\nonumber\\
  & & \frac{\partial w_{d} }{\partial \chi_{01}} = g_{3}M_{\chi}\chi_{1}+g_{4}(\varphi_{3}\eta_{2}-\varphi_{2}\eta_{1})=0              \nonumber\\
  & & \frac{\partial w_{d} }{\partial \chi_{02}} = g_{3}M_{\chi}\chi_{3}+g_{4}(\varphi_{2}\eta_{2}-\varphi_{1}\eta_{1})=0              \nonumber\\
  & & \frac{\partial w_{d} }{\partial \chi_{03}} = g_{3}M_{\chi}\chi_{2}+g_{4}(\varphi_{1}\eta_{2}-\varphi_{3}\eta_{1})=0              \nonumber\\
  & & \frac{\partial w_{d} }{\partial \eta_{0}} = h_{1}(\varphi^{2}_{1}+2\varphi_{2}\varphi_{3})+2h_{2}\eta_{1}\eta_{2}=0
\end{eqnarray}
There are two un-equivalent solutions for the equations, one solution gives
\begin{equation}\label{eq:Ve11}
    \qquad \langle \varphi \rangle = v_{\varphi}\left(
    \begin{array}{c}
    0 \\
    1 \\
    0 \\
    \end{array}
    \right)           , \quad
    \langle \eta \rangle=v_{\eta}\left(
    \begin{array}{c}
    0 \\
    1 \\
    \end{array}
    \right)            , \quad
    \langle \chi \rangle = v_{\chi}\left(
    \begin{array}{c}
    0 \\
    0 \\
    1 \\
    \end{array}
    \right)
\end{equation}
with
\begin{equation}\label{eq:condition Ve11}
   \qquad v_{\varphi}=-\frac{g_{2}}{2g_{1}}v_{\eta}, \quad
   v_{\chi}=-\frac{g_{2}g_{4}}{2g_{1}g_{3}M_{\chi}}v^{2}_{\eta}, \quad
v_{\eta}\ \text{undetermined}
\end{equation}
another solution of the form $\langle v_{\varphi} \rangle = (1,1,1)^{T}v_{\varphi}, \langle \chi \rangle = (1,1,1)^{T}v_{\chi},
\langle \eta \rangle=(1,-1)^{T}v_{\eta}$ is forbidden by the last term in Eq.~\eqref{eq:wde11}.

The minimum equation of the two singlets $\sigma$ and $\vartheta$ is simply given as
\begin{eqnarray}\label{eq:wde12}
\qquad \frac{\partial w_{d} }{\partial \sigma_{0}} = q_{1}\sigma^{2}+q_{2}\vartheta^{2} = 0
\end{eqnarray}
The equation in~\eqref{eq:wde12} lead to the following non-trivial solution
\begin{equation}\label{eq:Ve12}
\qquad  \langle\sigma\rangle=v_{\sigma},\quad \langle\vartheta\rangle=v_{\vartheta},
\end{equation}
with
\begin{equation}\label{eq:condition Ve12}
\qquad  v^{2}_{\sigma}=-\frac{q_{2}}{q_{1}}v^{2}_{\vartheta}, \quad v_{\vartheta}\ \text{undetermined}
\end{equation}

The VEVs of the above flavon fields in $\Phi^{e}$ will mainly determine the diagonal elements of mass matrices of charged fermions at LO, while
the other two flavons $\xi$ and $\rho$ in $\Phi^{e}$ appear in off-diagonal entries at LO. Their vacuum configurations are easily obtained
by the minimization equations
\begin{eqnarray}\label{eq:wde2}
& &  \frac{\partial w_{d} }{\partial \xi_{01}} = 2r_{1}(\xi^{2}_{1}-\xi_{2}\xi_{3})+r_{2}\xi_{1}\rho = 0 \nonumber\\
& &  \frac{\partial w_{d} }{\partial \xi_{02}} = 2r_{1}(\xi^{2}_{2}-\xi_{1}\xi_{3})+r_{2}\xi_{3}\rho = 0 \nonumber\\
& &  \frac{\partial w_{d} }{\partial \xi_{03}} = 2r_{1}(\xi^{2}_{3}-\xi_{1}\xi_{2})+r_{2}\xi_{2}\rho = 0 
\end{eqnarray}
The equations in~\eqref{eq:wde2} lead to three sets of un-equivalent solutions. The first set is
\begin{equation}\label{eq:Ve2}
  \qquad \langle \xi \rangle=v_{\xi}\left(
                                      \begin{array}{c}
                                        1 \\
                                        0 \\
                                        0 \\
                                      \end{array}
                                    \right)
  , \quad  \langle \rho \rangle = v_{\rho},
\end{equation}
with
\begin{equation}\label{eq:condition Ve2}
  \qquad v_{\xi}=-\frac{r_{2}}{2r_{1}}v_{\rho},
  \qquad v_{\rho}\ \text{undetermined}
\end{equation}
The second non-trivial solution is
\begin{equation}\label{eq:Ve22}
  \qquad \langle \xi \rangle=v_{\xi}\left(
                                      \begin{array}{c}
                                        0 \\
                                        0 \\
                                        0 \\
                                      \end{array}
                                    \right)
  , \quad  \langle \rho \rangle = v_{\rho},
\end{equation}
with $v_{\rho}$ undetermined. The third solution is
\begin{equation}\label{eq:Ve23}
  \qquad \langle \xi \rangle=v_{\xi}\left(
                                      \begin{array}{c}
                                        1 \\
                                        1 \\
                                        1 \\
                                      \end{array}
                                    \right)
  , \quad  \langle \rho \rangle = 0,
\end{equation}
with $v_{\xi}$ undetermined. The first solution in Eq.~\eqref{eq:Ve2} is chosen as the valid solution in our model. As elucidated
in~\cite{Ding/2011}, the vacuum configurations of Eqs.~(\ref{eq:Ve11}),~(\ref{eq:Ve12}),~(\ref{eq:Ve2}) and~(\ref{eq:Vf1}) are not the unique
solutions of respective minimization equations. The other minima of the scalar potential can be derived by acting the original VEVs
Eqs.~(\ref{eq:Ve11}),~(\ref{eq:Ve12}),~(\ref{eq:Ve2}) and~(\ref{eq:Vf1}) with the elements of the flavor group $S_{4}$. Nevertheless,
the new minima which are equivalent to the original ones can not lead to different physical results, i.e., the resulting fermion masses
and mixing parameters are exactly the same as the original minima did. Hence without of generality the original vacuum alignments
are chosed in the model, and the other scenarios with different phases only are related by the field redefinitions. Besides the non-trivial
solutions, the trivial solutions with vanishing flavon VEVs cannot be excluded in principle, either. The problem could be solved by the
introduction of the soft mass terms with the form
$m^{2}_{\xi}|\xi|^{2}+m^{2}_{\rho}|\rho|^{2}+\tilde{m}^{2}_{\xi}\xi^{2}+\tilde{m}^{2}_{\rho}\rho^{2}$ for $\xi$ and $\rho$.
By taking the mass parameters $m^{2}_{\xi,\rho}$ and $\tilde{m}^{2}_{\xi,\rho}$ to be negative values, one can check that only
the configurations in Eqs.~(\ref{eq:Ve11}),~(\ref{eq:Ve12}),~(\ref{eq:Ve2}) and~(\ref{eq:Vf1}) can be the lowest minimum of scalar potential
and more stable than the vanishing ones.

The minimization equations for vacuum configurations of $\phi, \Delta$ and $\zeta$ are given as following
\begin{eqnarray}
& & \frac{\partial w_{d} }{\partial \phi_{01}} = f_{1}(\Delta_{1}\phi_{2}-\Delta_{2}\phi_{3}) = 0  \nonumber \\
& & \frac{\partial w_{d} }{\partial \phi_{02}} = f_{1}(\Delta_{1}\phi_{1}-\Delta_{2}\phi_{2}) = 0  \nonumber \\
& & \frac{\partial w_{d} }{\partial \phi_{03}} = f_{1}(\Delta_{1}\phi_{3}-\Delta_{2}\phi_{1}) = 0  \nonumber \\
& & \frac{\partial w_{d} }{\partial \Delta_{01}} = f_{2}(\phi^{2}_{3}+2\phi_{1}\phi_{2})+f_{3}\Delta^{2}_{1} = 0 \nonumber \\
& & \frac{\partial w_{d} }{\partial \Delta_{02}} = f_{2}(\phi^{2}_{2}+2\phi_{1}\phi_{3})+f_{3}\Delta^{2}_{2} = 0 \nonumber \\
& & \frac{\partial w_{d} }{\partial \zeta_{0}} = f_{4}\zeta\zeta+f_{5}(\phi^{2}_{1}+2\phi_{2}\phi_{3})+2f_{6}\Delta_{1}\Delta_{2} = 0
\end{eqnarray}
The solutions for above equations are listed as below
\begin{equation}\label{eq:Vf1}
    \qquad \langle\phi\rangle=v_{\phi}\left(
    \begin{array}{c}
    1 \\
    1 \\
    1 \\
    \end{array}
    \right), \quad
    \langle \Delta \rangle=v_{\Delta}\left(
    \begin{array}{c}
    1 \\
    1 \\
    \end{array}
    \right), \quad
    \langle \zeta \rangle=v_{\zeta},
\end{equation}
with the conditions
\begin{equation}\label{eq:condition Vf1}
     \qquad v^{2}_{\phi}=-\frac{f_{3}}{3f_{2}}v^{2}_{\Delta}, \quad
     v^{2}_{\Delta}=\frac{f_{2}f_{4}}{f_{3}f_{5}-2f_{2}f_{6}}v^{2}_{\zeta},  \quad
     v_{\zeta} \ \text{undetermined}
\end{equation}
The solution~\eqref{eq:Vf1} is used to produce the Tri-Bimaximal mixing pattern in the following sections.

The LO results in previous sections imply that the order of magnitude for $\langle\Phi^{e}\rangle$ and $\langle\Phi^{\nu}\rangle$
scaled by the cutoff $\Lambda$ are not common. To be specific the order of VEVs are restricted as follows
\begin{equation}\label{eq:hierarchy VeVf}
     \qquad \frac{v_{\varphi}}{\Lambda} \sim \frac{v_{\eta}}{\Lambda} \sim \frac{v_{\chi}}{\Lambda}
     \sim \frac{v_{\sigma}}{\Lambda} \sim \frac{v_{\vartheta}}{\Lambda} \sim \frac{v_{\xi}}{\Lambda} \sim \frac{v_{\rho}}{\Lambda}
     \sim \lambda^{2}_{c}, \qquad
     \frac{v_{\phi}}{\Lambda} \sim \frac{v_{\Delta}}{\Lambda} \sim \frac{v_{\zeta}}{\Lambda} \sim\lambda^{3}_{c}
\end{equation}
It is natural to require the subleading corrections to $\langle\Phi^{e}\rangle$ should be smaller than
$m_{\mu}/m_{\tau}\sim \mathcal{O}(\lambda^{2}_{c})$, or even more strictly smaller than $m_{e}/m_{\tau}\sim \mathcal{O}(\lambda^{3}_{c})$.
Because of the constrain of the auxiliary Abelian shaping symmetries $Z_{4}\times Z_{6} \times Z_{5}\times Z_{2}$, the subleading corrections
to superpotential $w^{e}_{d}$ and $w^{\nu}_{d}$ are suppressed by $1/\Lambda$ and $1/\Lambda^{2}$, respectively, see \hyperref[B]{APPENDIX B}
for detail. 

Next we shall briefly discuss the SU(5) GUT breaking scenario in the present scheme. The Higgs sector is composed of $H_{5}$, $H_{\bar{5}}$, $H_{45}$,
$H_{\overline{45}}$,  and the gauge group is broken by the VEV of the adjoint field $H_{24}$. The LO invariant interactions between
the Higgs superfields and the adjoint field under the flavor symmetries are given by
\begin{eqnarray}\label{eq:wH}
  & & w_{H}=\frac{f_{H1}}{\Lambda}H_{24}H_{24}\sigma\rho+\sum^{5}_{i=1}\frac{f'_{Hi}}{\Lambda^{3}}H_{24}H_{24}\mathcal{O}^{H}_{i}
  +\frac{g_{H1}}{\Lambda^{4}}H_{\bar{5}}H_{5}H_{24}(\chi\chi)_{2}(\chi\phi)_{2} \nonumber\\
  & &\qquad +\frac{g_{H2}}{\Lambda^{4}}H_{\bar{5}}H_{5}H_{24}(\chi\chi)_{3_{1}}(\chi\phi)_{3_{1}}
   +\frac{g_{H3}}{\Lambda^{4}}H_{\bar{5}}H_{5}H_{24}(\chi\chi)_{3_{1}}(\chi\Delta)_{3_{1}},
\end{eqnarray}
in which the operators $\mathcal{O}^{H}_{i}$ include the following contractions of $S_{4}$
\begin{equation}\label{eq:OHi}
\qquad  \mathcal{O}^{H}_{i}=\{\sigma\rho\zeta\zeta,\sigma\rho(\Delta\Delta)_{1_{1}},\sigma\rho(\phi\phi)_{1_{1}},
\sigma\xi(\phi\phi)_{3_{1}},\sigma\xi(\phi\Delta)_{3_{1}}\}
\end{equation}
Plugging the VEVs of the flavons appear in the above equation, we can obtain the following
\begin{equation}\label{eq:wH2}
\qquad w_{H}=f_{H}H_{24}H_{24}+g_{H}H_{\bar{5}}H_{5}H_{24},
\end{equation}
with the coefficients
\begin{eqnarray}
& & f_{H}=(f_{H_{1}}+f'_{H_{1}}\frac{v^{2}_{\zeta}}{\Lambda^{2}}+2f'_{H_{2}}\frac{v^{2}_{\Delta}}{\Lambda^{2}}
+3f'_{H_{3}}\frac{v^{2}_{\phi}}{\Lambda^{2}})\frac{v_{\sigma}v_{\rho}}{\Lambda}
+2f'_{H_{5}}\frac{v_{\sigma}v_{\xi}v_{\phi}v_{\Delta}}{\Lambda^{3}}, \nonumber\\
& & g_{H}=(g_{H1}-2g_{H2})\frac{v^{3}_{\chi}v_{\phi}}{\Lambda^{4}}
\end{eqnarray}
The superpotential in equation~\eqref{eq:wH} breaks $U(1)_{R}$ due to all the fields involved carrying 0 unit $U(1)_{R}$ charge. Even taking into
account the operators with the driving fields, Higgs fields and flavon fields combined together, the vanishing VEVs of the driving fields signify
the operators have no contributions to the scalar potential. In order to completely understand the GUT symmetry breaking, one may consider the
ultraviolet (UV) completion of the effective model. By adding the fields with non-zero unit $U(1)_{R}$ charge, a $U(1)_{R}$ conserving superpotential
can give rise to the terms in Eq.~\eqref{eq:wH}. For the examples of the ultraviolet completion in realistic flavor models, please
see Refs.~\cite{Gehrlein/2015gA5,Bjorkeroth/2015gA4}.

\section{Corrections}\label{S5}
The subleading corrections to superpotentials above arise from the higher dimensional operators which are suppressed by at least one
power of $1/\Lambda$ and constrained by the symmetries. In this work the modified superpotential of driving fields $w_{d}$ would give
rise to the shifts of LO VEVs in $\Phi^{e}$ and $\Phi^{\nu}$. Fermions' correctional masses and mixing matrices are obtained by adding
the higher order operators and the shifted vacua of flavon fields. The combinations $\phi\phi$, $\phi\Delta$, $\Delta\Delta$ and $\zeta\zeta$
are invariant under the auxiliary $\mathcal{G}_{A}=Z_{4}\times Z_{6}\times Z_{5}\times Z_{2}$, hence it is always viable to add these
combinations with any power on the top of each LO terms. It is expected that the subleading corrections are not impossible to destroy
the LO predictions. Even though the expectation maybe correct, we would also like to present the detailed analysis of the subleading corrections.

\subsection{Corrections to Vacuum Alignment}\label{S5sub1}
Here we just present the final results of shifted VEVs. The detailed calculation procedure is presented in \hyperref[B]{APPENDIX B}. The
modified VEVs of all flavons in $\Phi^{e}$ are of the form as follows
\begin{eqnarray}\label{eq:subVe}
  & &\langle \varphi \rangle = \left(
      \begin{array}{c}
      \delta v_{\varphi_{1}} \\
      v_{\varphi}+\delta v_{\varphi_{2}} \\
      \delta v_{\varphi_{3}} \\
      \end{array}
      \right), \quad
      \langle \eta \rangle = \left(
      \begin{array}{c}
      \delta v_{\eta_{1}} \\
      v_{\eta} \\
      \end{array}
      \right), \quad
      \langle \chi \rangle = \left(
      \begin{array}{c}
      \delta v_{\chi_{1}} \\
      \delta v_{\chi_{2}} \\
      v_{\chi}+\delta v_{\chi_{3}} \\
      \end{array}
      \right), \quad
      \langle \xi \rangle = \left(
      \begin{array}{c}
      v_{\xi}+\delta v_{\xi_{1}} \\
      \delta v_{\xi_{2}} \\
      \delta v_{\xi_{3}} \\
      \end{array}
      \right),\nonumber\\
  & & \langle \rho \rangle = v_{\rho},
     \quad\langle \sigma \rangle = v_{\sigma}+\delta v_{\sigma},
     \quad  \langle \vartheta \rangle = v_{\vartheta},
\end{eqnarray}
with $v_{\eta}$, $v_{\vartheta}$ and $v_{\rho}$undetermined. We remark that the correctional results of each component of the flavons in $\Phi^{e}$
are different. Similarly the shifted vacua of the scalars in $\Phi^{\nu}$ read
\begin{equation}\label{eq:subVf1}
\qquad \langle \phi \rangle=\left(
\begin{array}{c}
v_{\phi}+\delta v_{\phi_{1}} \\
v_{\phi}+\delta v_{\phi_{2}} \\
v_{\phi}+\delta v_{\phi_{3}} \\
\end{array}
\right), \quad
\langle \Delta \rangle = \left(
\begin{array}{c}
v_{\Delta}+\delta v_{\Delta_{1}} \\
v_{\Delta}+\delta v_{\Delta_{2}} \\
\end{array}
\right), \quad
\langle \zeta \rangle = v_{\zeta},
\end{equation}
where $v_{\zeta}$ is still undetermined. We also remark that in fact the correctional results of each components of the flavons in
$\Phi^{\nu}$ are exactly the same, it is an important feature of the model which makes the Tri-Bimaximal mixing pattern still holds at subleading
corrections in neutrino sector. Take into account the hierarchical VEVs $\langle\Phi^{e}\rangle/\Lambda \sim \lambda^{2}_{c}$
and $\langle\Phi^{\nu}\rangle/\Lambda \sim\lambda^{3}_{c}$, and the subleading operators linear in driving fields are suppressed by
different power of $1/\Lambda$, the shifts also have different order of magnitude
\begin{eqnarray}\label{eq:shift vacua order}
& &  \frac{\delta v_{\varphi_{i}}}{v_{\varphi}} \sim
 \frac{\delta v_{\eta_{1}}}{v_{\eta}} \sim
 \frac{\delta v_{\chi_{i}}}{v_{\chi}} \sim \lambda^{4}_{c},
\quad  \frac{\delta v_{\sigma}}{v_{\sigma}} \sim \lambda^{3}_{c},
     \quad \frac{\delta v_{\xi_{1,2}}}{v_{\xi}} =  0,
     \quad \frac{\delta v_{\xi_{3}}}{v_{\xi}} \sim \lambda^{2}_{c},
\quad  \frac{\delta v_{\phi_{i}}}{v_{\phi}} \sim
 \frac{\delta v_{\Delta_{i}}}{v_{\Delta}}\sim \lambda^{6}_{c}
\end{eqnarray}
\subsection{Corrections to neutrino}\label{S5sub2}
Due to the auxiliary symmetry $Z_{4}\times Z_{6} \times Z_{5}\times Z_{2}$, the subleading corrections to neutrino Majorana mass matrix only appear
at next to next to leading order (NNLO). The higher dimensional operators arising from inserting bilinear invariant combinations of
$\phi$, $\Delta$ and $\zeta$ into the superfield $N^{c}N^{c}$ in all possible ways, the corresponding terms are expressed explicitly as follows:
\begin{eqnarray}\label{eq:subMajorana}
& &
\frac{z_{c1}}{\Lambda}N^{c}N^{c}(\phi\phi)_{1_{1}}+\frac{z_{c2}}{\Lambda}N^{c}N^{c}(\phi\phi)_{2}+\frac{z_{c3}}{\Lambda}N^{c}N^{c}(\phi\phi)_{3_{1}}
  +\frac{z_{c4}}{\Lambda}N^{c}N^{c}(\phi\Delta)_{3_{1}}                            \nonumber\\
& & +\frac{z_{c5}}{\Lambda}N^{c}N^{c}(\Delta\Delta)_{1_{1}}+\frac{z_{c6}}{\Lambda}N^{c}N^{c}(\Delta\Delta)_{2}
  +\frac{z_{c7}}{\Lambda}N^{c}N^{c}\zeta\zeta
\end{eqnarray}
Inserting the LO VEVs of the flavons in $\Phi^{\nu}$, the NNLO corrections to Majorana mass matrix will be
\begin{eqnarray}\label{eq:subMajorana mass matrix}
& &  \delta M_{M} = \left(
  \begin{array}{ccc}
  1 & 0 & 0 \\
  0 & 0 & 1 \\
  0 & 1 & 0 \\
  \end{array}
  \right)\frac{3z_{c1}v^{2}_{\phi}+2z_{c5}v^{2}_{\Delta}+z_{c7}v^{2}_{\zeta}}{\Lambda}
  +\left(
     \begin{array}{ccc}
       0 & 1 & 1 \\
       1 & 1 & 0 \\
       1 & 0 & 1 \\
     \end{array}
   \right)\frac{3z_{c2}v^{2}_{\phi}+z_{c6}v^{2}_{\Delta}}{\Lambda} \nonumber\\
& & \qquad \qquad  +\left(
     \begin{array}{ccc}
       2 & -1 & -1 \\
       -1 & 2 & -1 \\
       -1 & -1 & 2 \\
     \end{array}
   \right)\frac{2z_{c4}v_{\phi}v_{\Delta}}{\Lambda}
\end{eqnarray}
We can note the form of $\delta M_{M}$ is still compatible with TBM mixing. Denote the three terms that following the matrices as
$S_{1},S_{2},S_{3}$ respectively, the eigenvalues of the subleading correctional Majorana masses are easily obtained as follows
\begin{eqnarray}\label{eq:subMajorana mass}
  \qquad dM_{1}= S_{1}-S_{2}+3S_{3},
  \qquad dM_{2}=S_{1}+2S_{2},
  \qquad dM_{3}=-S_{1}+S_{2}+3S_{3}
\end{eqnarray}
The possible content that spoils the TBM mixing only arising from Dirac mass terms, whose corresponding subleading superpotential is comprised of the
shifted VEVs at LO and higher dimensional operators as following
\begin{equation}\label{eq:subDirac}
\qquad \delta w_{D} = \sum_{i=1}^{2}\frac{y_{\nu_{i}}}{\Lambda}FN^{c}\delta\Phi_{i}^{\nu} H_{5}
          +\sum_{i=1}^{12}\frac{y'_{\nu_{i}}}{\Lambda^{3}}(FN^{c})_{c_{i}}(\Phi^{\nu}\Phi^{\nu}\Phi^{\nu})_{c_{i}} H_{5}
\end{equation}
in which the $\delta\Phi^{\nu}_{i}$ denotes the shifted vacua of flavon $\Phi^{\nu}_{i}$, and $c_{i}$ indicate all possible $S_{4}$
contractions. Substituting the unique LO vacua structures of $\Phi^{\nu}$ in Eq.~\eqref{eq:Vf1} and the shifted VEVs of $\delta\Phi^{\nu}$
in Eq.~\eqref{eq:delta Vf1} into Eq.~\eqref{eq:subDirac}, one can check that actually the two sets of superpotential invariants still maintain TBM
mixing after symmetry breaking, and the modified Dirac mass terms are exactly the same as the mass structure in Eq.~\eqref{eq:LO Dirac mass} at LO
and in Eq.~\eqref{eq:subMajorana mass matrix} at NNLO, thus the corrections could be absorbed into Yukawa couplings by redefining
parameters $y_{\nu_{i}}$ in the LO part. Note that the contraction $1_{1}$ of $S_{4}$ also induces extra term in Dirac mass matrix, which arise
from the second term in above Eq.~\eqref{eq:subDirac}, i.e., the terms with $c_{i}=1_{1}$, however, can not be absorbed into the redefinition
of parameters $y_{\nu_{i}}$. To be specific the correctional extra Dirac mass matrix is found to be
\begin{equation}\label{eq:extraDiracMatrix}
\delta m_{D}=\left(
  \begin{array}{ccc}
    1 & 0 & 0 \\
    0 & 0 & 1 \\
    0 & 1 & 0 \\
  \end{array}
\right)\frac{v_{5}}{\Lambda}c,\quad
  c=(6y'_{\nu_{1}}v^{2}_{\phi}v_{\Delta}+2y'_{\nu_{2}}v^{3}_{\Delta})/\Lambda^{2}.
\end{equation}
Collected all the Dirac mass matrices in Eq.~\eqref{eq:LO Dirac mass} and Eq.~\eqref{eq:extraDiracMatrix}, the structure is exact the same with
Eq.~\eqref{eq:subMajorana mass matrix}, and can be still diagonalized by TBM mixing matrix. The stability of TBM mixing in neutrino sector is
guaranteed by the stable VEV structures of $\Phi^{\nu}$, see \hyperref[B]{APPENDIX B}. It is a salient feature of the model that Tri-Bimaximal mixing
still holds even at NNLO.

Take into account these subleading contributions to both Dirac and Majorana mass matrices, we can easily rewrite the modified light neutrino masses
as follows
\begin{eqnarray}\label{eq:sub light mass}
  \qquad m_{1} = \Big|-\frac{(3a-b+c)^{2}}{M+dM_{1}}\Big|\frac{v^{2}_{5}}{\Lambda^{2}},
  \qquad m_{2} = \Big|-\frac{(2b+c)^{2}}{M+dM_{2}}\Big|\frac{v^{2}_{5}}{\Lambda^{2}},
  \qquad m_{3} = \Big|\frac{(3a+b-c)^{2}}{M-dM_{3}}\Big|\frac{v^{2}_{5}}{\Lambda^{2}}
\end{eqnarray}
Note that $a$, $b$ in above expressions are not exactly the same as in Eq.~\eqref{eq:mlight}, because of the redefinition of LO parameters
$y_{\nu_{i}}$ by absorbing the corrections to Dirac masses in Eq.~\eqref{eq:subDirac}. However the subleading corrections are too small to change the
order of magnitudes, we can treat them unchanged. The extra term $c$ is also extremely small compared with $a$ and $b$, which can not affect the
magnitude of neutrino mass as well.

\subsection{Corrections to Charged Fermions}\label{S5sub3}
There are two sources for the correctional contributions to charged fermions, one is the higher-dimensional operators, another is the shifted VEVs at
LO. The masses and mixings for charged fermions have been well determined at LO, hence the subleading effects are expected to be negligible.
For sake of simplicity, we shall drop the detailed procedure of calculations, and focus on the generical results of subleading effects.
Ignoring the $\mathcal{O}(1)$ couplings and setting all cutoff to be $\Lambda$, the higher dimensional invariant operators
under all the symmetries are generically of the form
\begin{equation}\label{eq:subCF}
  \qquad \delta w_{U}\propto T_{i}T_{j}\Big(\frac{\Phi^{e}}{\Lambda}\Big)^{m}\Big(\frac{\Phi^{\nu}}{\Lambda}\Big)^{n}H_{5,45},
  \quad \delta w_{D}\propto T_{i}F\Big(\frac{\Phi^{e}}{\Lambda}\Big)^{p}\Big(\frac{\Phi^{\nu}}{\Lambda}\Big)^{q}H_{\bar{5},\overline{45}}
  +T_{i}FH_{24}H_{\bar{5},\overline{45}}\Big(\frac{\Phi^{e}}{\Lambda}\Big)^{r}\Big(\frac{\Phi^{\nu}}{\Lambda}\Big)^{s}
\end{equation}
with $m+n\geqslant4$, $p+q\geqslant4$ and $r+s\geqslant4$. The order of $\langle\Phi^{e}\rangle$ and $\langle\Phi^{\nu}\rangle$ imply that
the subleading contributions to the entries of $M_{U}$ and $M_{D}$ from $\delta w_{U}$ and the first term of $\delta w_{D}$ are of order
$\epsilon^{4}$ or less. Depending on the ways SU(5) indices contract, the largest corrections of order $\epsilon^{2}\delta$ to second column
of $M_{D}$ arise from the operators $[FH_{\bar{5}}]_{\textbf{15}}[T_{2}H_{24}]_{\overline{\textbf{15}}}\sigma(\xi\phi,\xi\Delta)_{3_{1}}$.
Also the operators give rise to vanishing contributions to $M_{\ell}$ due to $y_{e}/y_{d}=0$~\cite{Antusch/2014GUT}.
In short the higher-dimensional operators contribute negligible effect
with respect to the LO results in Section~\ref{S3}.

Plugging the shifted vacua $\delta v_{\Phi}$ in~Eq.~\eqref{eq:shift vacua order} into the LO superpotential in section~\ref{S3}, one may check that 
the correctional mass entries are much smaller than respective LO ones. All the corrections have no significant impact on the LO masses, 
$m_{u,c,t}$, $m_{d,s,b}$ and $m_{e,\mu,\tau}$, and mixings $V_{CKM}$ and $U_{PMNS}$. In conclusion the LO predictions can not be spoiled by
the two types of subleading effects.

\section{Phenomenology}\label{S6}
The model we have constructed has only analytic form despite of some parameters within it. In the next step we shall present the phenomenological
numerical results of some observables we interested. We require the input oscillation parameters $\Delta m^{2}_{sol}$, $\Delta m^{2}_{atm}$,
$\sin^{2}\theta_{12}$, $\sin^{2}\theta_{13}$, and $\sin^{2}\theta_{23}$ to lie in their $3\sigma$ intervals which are taken from
Ref.~\cite{Forero/2014FIT}. It is easy to express the light neutrino mass spectrum with well determined mass differences and the lightest neutrino mass
$m_{\textrm{lightest}}=m_{1}\;(m_{3})$ in NH (IH) spectrum as follows
\begin{eqnarray}\label{eq:mass hierarchy}
& & \textrm{NH}: m_{1} < m_{2} < m_{3},\quad m_{2}=\sqrt{\Delta m^{2}_{sol}+m^{2}_{1}}, \quad m_{3}=\sqrt{\Delta m^{2}_{atm}+m^{2}_{1}} \nonumber\\
& & \textrm{IH}:\;  m_{3} < m_{1} < m_{2},\quad m_{1}=\sqrt{m^{2}_{3}-\Delta m^{2}_{atm}}, \quad m_{2}=\sqrt{m^{2}_{3}-\Delta m^{2}_{atm}
+\Delta m^{2}_{sol}}
\end{eqnarray}
\subsection{Mixing angles}\label{S6sub1}
\begin{figure}[htb]
  \centering
  \subfigure[]{
  \label{fig:subfig:12} 
  \includegraphics[width=2.6in]{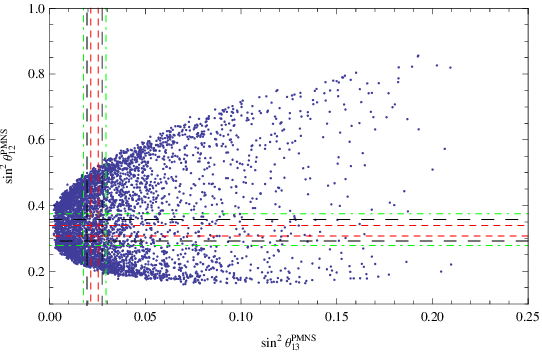}}
  \hspace{0.02in}
  \subfigure[]{
  \label{fig:subfig:23NH} 
  \includegraphics[width=2.6in]{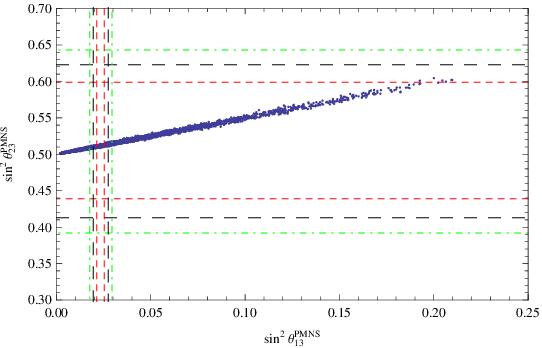}}
    \caption{\small The allowed region of $\sin^{2}\theta^{PMNS}_{12}-\sin^{2}\theta^{PMNS}_{13}$ (left panel)
    and $\sin^{2}\theta^{PMNS}_{23}-\sin^{2}\theta^{PMNS}_{13}$
    (right panel).}
  \label{fig:theta} 
\end{figure}

In the numerical analysis all the coupling coefficients in analytic expressions, i.e., Eq.~\eqref{eq:LO angles}, of mixing angles are
taken to random complex numbers with their absolute values (or say modulus) within an interval [1/2, 3/2], and the small parameters $\epsilon$ and
$\delta$ can be fixed at 0.05 and 0.01 respectively as demonstration values. The ratio $v_{\overline{45}}/v_{\bar{5}}$ is chosen the typical
value 0.22. The analytic expressions of leptonic mixing angles are shown in Eq.~\eqref{eq:LO angles}, which include the deviations from TBM
mixing values. Thus we can estimate the allowed region of mixing parameters
numerically. The allowed regions of $\sin^{2}\theta^{PMNS}_{13}-\sin^{2}\theta^{PMNS}_{12}$ and $\sin^{2}\theta^{PMNS}_{13}-\sin^{2}\theta^{PMNS}_{23}$
are shown in Fig.~\ref{fig:subfig:12} and Fig.~\ref{fig:subfig:23NH} respectively. The horizontal lines show the
$3\sigma$ (green), $2\sigma$ (black) and $1\sigma$ (red) boundaries of the mixing angles $\theta^{PMNS}_{12}$ and $\theta^{PMNS}_{23}$, while the
vertical line presents the corresponding boundaries of $\theta^{PMNS}_{13}$.

\begin{figure}[htb]
  \centering{
  \includegraphics[width=2.6in]{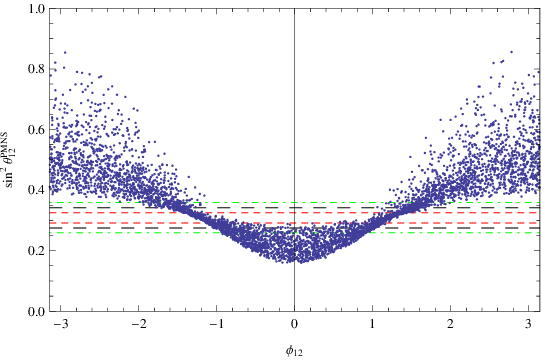}}
  \caption{\small Correlation of $\sin^{2}\theta^{PMNS}_{12}$ and $\phi_{12}=\arg(y^{d}_{12}/y^{d}_{22}$).}\label{fig:phase}
\end{figure}

As showed in Eq.~\eqref{eq:12phi}, the deviation of $\theta^{PMNS}_{12}$ from its TBM value $\theta^{\nu}_{12}$ is mainly controlled by the complex
phase, defined as $\phi_{12}$, of $y^{d}_{12}/y^{d}_{22}$ in which we have restricted the module of $y^{d}_{12}/y^{d}_{22}$ in the interval [1/3, 3].
Thus we can estimate the effect of the phase on mixing angle $\theta^{PMNS}_{12}$, see Fig.~\ref{fig:phase}. The horizontal lines stand for the
confidence level as in Fig.~\ref{fig:subfig:12}, and it is obviously that $\phi_{12}\sim \pi/2$ or $-\pi/2$ are favoured so that $\theta^{PMNS}_{12}$
can lie in the $3\sigma$ interval in the present model.
\subsection{Sum of neutrino masses, Neutrinoless double beta decay}\label{S6sub2}
\begin{figure}[htb]
  \centering
  \subfigure[]{
  \label{fig:subfig:mratio} 
  \includegraphics[width=2.7in]{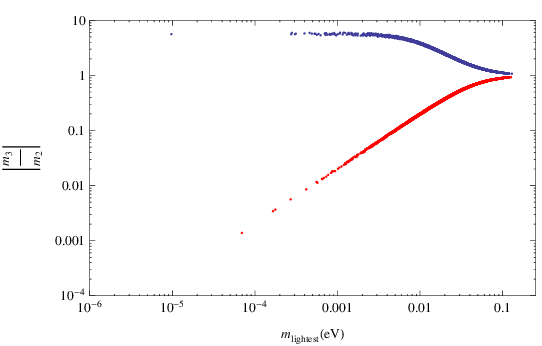}}
  \hspace{0.1in}
  \subfigure[]{
  \label{fig:subfig:sum of masses} 
  \includegraphics[width=2.6in]{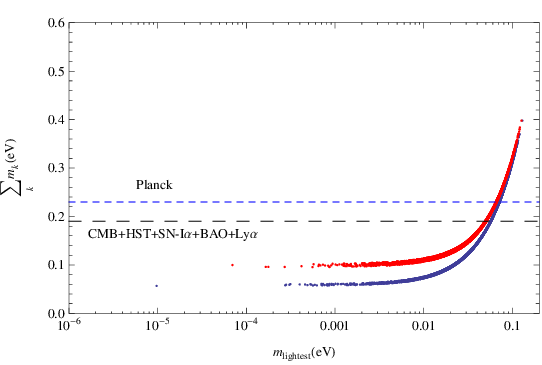}}
  \caption{\small Scatter plots of $|m_{3}/m_{2}|$ (left panel) and sum of neutrino masses $\sum_{k}m_{k}$ (right panel) as
  the function of the lightest neutrino mass $m_{\textrm{lightest}}$. In both plots blue corresponds to NH mass spectrum and red to IH mass spectrum.}
  \label{mass1} 
\end{figure}

First we consider two simple arithmetical relationship between light neutrino masses: the ratio $|m_{3}/m_{2}|$ and sum of all masses against
the lightest neutrino mass. The plot of the ratio $|m_{3}/m_{2}|$ and sum of light neutrino mass $\sum_{k}m_{k}$ as function of the lightest
neutrino mass $m_{\textrm{lightest}}$, which is $m_{1}$ ($m_{3}$) for NH (IH) mass spectrum, are shown in Fig.~\ref{fig:subfig:mratio} and
Fig.~\ref{fig:subfig:sum of masses}, respectively. Note that the horizontal lines in Fig.~\ref{fig:subfig:sum of masses} represents the
cosmological bound at 0.19 eV (black), corresponding to the combined observational data from~\cite{cosm}, and the upper bounds 0.23 eV from
Planck~\cite{Planck/2013}. The ratio tends to a degenerate mass spectrum in both cases as the value of $m_{\textrm{lightest}}\rightarrow$ 0.1eV which
is disfavoured in the model. The masses sum $\sum_{k}m_{k}$ in the model is predicted too similar for both hierarchical mass spectrums to
be distinguished using the cosmological bound on the sum of neutrino masses.

\begin{figure}[htb]
  \centering
  \subfigure[]{
  \label{fig:subfig:mee} 
  \includegraphics[width=2.6in]{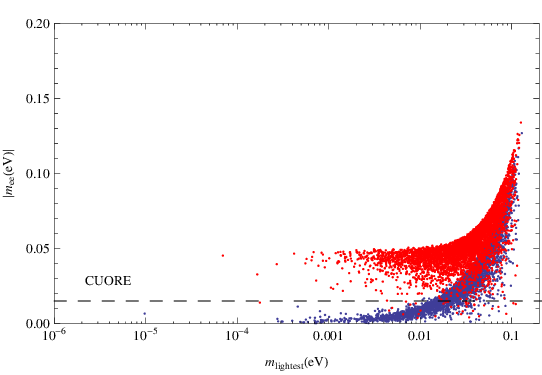}}
  \hspace{0.1in}
  \subfigure[]{
  \label{fig:subfig:mbeta} 
  \includegraphics[width=2.6in]{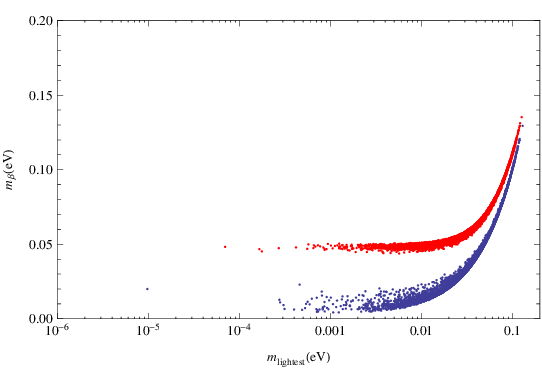}}
  \caption{\small Effective Majorana mass of $0\nu\beta\beta$ decay $|m_{ee}|$ (left panel) and effective mass of beta decay (right panel)
  $m_{\beta}$ as the function of the lightest neutrino mass $m_{\textrm{lightest}}$. In both plots blue corresponds to NH mass spectrum and red to
  IH mass spectrum.}
  \label{mass2} 
\end{figure}

The effective Majorana mass $|m_{ee}|$ determines the $0\nu\beta\beta$ decay amplitude, which is also the (11) element of neutrino mass matrix in
flavor basis, and meant a real and diagonal matrix of charged lepton mass. It is defined as follows
\begin{equation}\label{eq:mee}
 \qquad  |m_{ee}|=\Big|\sum_{i}(U_{PMNS})^{2}_{ei}m_{i}\Big|
\end{equation}
and the $\beta$ decay effective mass which could measure the non-zero neutrino masses is
\begin{equation}\label{eq:mbeta}
 \qquad  m_{\beta}=\Big[\sum_{k}\Big|(U_{PMNS})_{ek}\Big|^{2}m^{2}_{k}\Big]^{1/2}
\end{equation}
In the numerical analysis the mass differences are well-known input parameters, as explained in the beginning of this section, and the analytic form
of $m_{\textrm{lightest}}$ ($m_{1}$ or $m_{3}$) in Eq.~\eqref{eq:sub light mass} are the single variant of physical quantities $|m_{ee}|$ and
$m_{\beta}$. The predicted allowed region of the two effective masses against lightest mass, are shown in Fig.~\ref{fig:subfig:mee} and
Fig.~\ref{fig:subfig:mbeta}. The dashed line in Fig.~\ref{fig:subfig:mee} shows the future sensitivity of CUORE~\cite{CUORE/2008} experiment at 15 meV.
In the model $|m_{ee}|$ is predicted to below the upper limit of $|m_{ee}|$, which is constrained from the Heidelberg-Moscow experiment~\cite{HM/1999}.
As the $m_{\textrm{lightest}}$ grows to be around 0.1 eV, $|m_{ee}|$ tends to degenerate in both NH and IH mass spectrum.

The $\beta$ decay effective mass $m_{\beta}$ is predicted to be below the future sensitivity 0.2eV of the KATRIN~\cite{KATRIN/2001} experiment,
as showed in Fig.~\ref{fig:subfig:mbeta}. The vertical line represents the sensitivity.

\section{Conclusion}\label{S7}
In this paper, we have proposed a flavor model in the framework of SUSY SU(5) GUT based on $S_{4}\times Z_{4} \times Z_{6}\times Z_{5}\times Z_{2}$
flavor symmetry. In the model the first and third generations of \textbf{10} dimensional representation in SU(5) are all assigned to be $1_{1}$
of $S_{4}$. The second generation of \textbf{10} dimensional is to be $1_{2}$ of $S_{4}$. Right-handed neutrinos of singlet \textbf{1} in SU(5) and
three generations of $\bar{\textbf{5}}$ in SU(5) are all assigned to be $3_{1}$ of $S_{4}$. The flavons in $\Phi$ are divided into $\Phi^{e}$ in
charged fermion sector and $\Phi^{\nu}$ in neutrino sector, whose VEVs are of different orders of magnitude:
$\langle\Phi^{e}\rangle/\Lambda \sim \lambda^{2}_{c}$ and $\langle\Phi^{\nu}\rangle/\Lambda \sim \lambda^{3}_{c}$, with $\lambda_{c}$ being Cabibbo
angle. Also the energy scale $\Lambda$ is below the GUT scale.

The three right-handed neutrinos are SU(5) singlets, and the light neutrino masses are generated via type-I seesaw mechanism only. The diagonalization
of neutrino mass matrix leads to Tri-Bimaximal mixing pattern at LO, both normal and inverted hierarchy mass spectrums are allowed. The subleading
corrections to both Majorana and Dirac masses arise from the higher dimensional operators and shifted VEVs lead to the same mass structures as that at
LO, and give no change to the TBM mixing pattern. It is a salient feature of the present model that TBM mixing still holds exactly at NNLO.

The mass hierarchies of up-type quarks are controlled by the spontaneously symmetry broken. The top quark obtains its mass purely at tree level, the
LO mass of charm quark is derived only by two flavons $\sigma$ and $\vartheta$ in $\Phi^{e}$, while the LO mass of up quark is controlled by three
flavons: $\eta$ in $\Phi^{e}$ and $\Delta, \zeta$ in $\Phi^{\nu}$. Due to the moderate hierarchy assumptions
$\langle\Phi^{e}\rangle\sim\lambda^{2}_{c}\Lambda$ and $\langle\Phi^{\nu}\rangle\sim\lambda^{3}_{c}\Lambda$, the phenomenological favored mass
hierarchies of all three generations of up-type quarks are obtained, i.e., $m_{u}:m_{c}:m_{t}=\lambda^{8}_{c}:\lambda^{4}_{c}:1$. The mixing at LO only
exists between the last two generations, give rise to the mixing angle $\theta^{u}_{23}\sim \lambda^{2}_{c}$.

The mass texture of down-type quarks is similar to that of charged-leptons due to the same set of GUT operators, despite of the different
CG factors and transposed relations. The model predicts that bottom-tau unification $m_{b}=m_{\tau}$ as well as the popular Georgi-Jarlskog
relation $m_{\mu}=3m_{s}$. In addition the model also gives a new mass relation between electron and down quark, namely $m_{e}=\frac{8}{27}m_{d}$.
The new ratio arises from the splitted masses of heavy messenger fields by a specific novel CG factors from GUT symmetry breaking.
Concretely the non-singlet field is an adjoint representation, $H_{24}$, of SU(5) in our model. The novel CG factors caused by
$\langle H_{24}\rangle$ enter inversely in the desired Yukawa textures of quark-lepton, leading to the new mass relation. The model also
gives the CKM quark mixing matrix and a CKM-like mixing matrix of charged leptons. The resulting Cabibbo angle $\theta_{c}$ between
the first two families of down-quarks, together with the mixing angle between the first and
third generations all require a fine tuning $v_{\overline{45}}/v_{\bar{5}}\sim\lambda_{c}$. Combined the mixing angle $\theta^{u}_{23}$, all the
elements of CKM matrix can be derived properly. On the other hand the CKM-like mixing matrix of charged leptons also implies a sizable mixing
between electron and muon, given that $\theta^{e}_{12}\simeq\lambda_{c}$. The mixing angle is constrained by the
GUT relation $\theta^{e}_{12}=|\frac{c_{c}}{c_{b}}|\frac{m_{e}}{m_{\mu}}\frac{1}{\theta^{d}_{12}}$, only with the condition that
$|\frac{c_{c}}{c_{b}}|\sim\mathcal{O}(2/\lambda_{c})$ and $\theta^{d}_{12}\simeq\lambda_{c}$, the angle $\theta^{e}_{12}$ can be $\lambda_{c}$ as well.
The model with the novel CG factors indeed leads to $\frac{c_{c}}{c_{b}}=\frac{81}{8}$, which satisfies the condition. Finally the CKM-like mixing
matrix of charged leptons with $\theta^{e}_{12}\simeq\lambda_{c}$ modifies the vanishing $\theta^{\nu}_{13}$ in TBM mixing pattern to a sizable lepton
mixing angle $\theta^{PMNS}_{13}\simeq\lambda_{c}/\sqrt{2}$, in well compatible with experimental results.

We also present the subleading corrections to flavon alignment in detail, and we find that all VEVs $\langle\Phi^{\nu}\rangle$ actually receive very
small shifts along the same directions of the LO alignment even at NNLO corrections, but it is not the case for $\Phi^{e}$. The stable solutions of
$\langle\Phi^{\nu}\rangle$ also guarantee the stability of TBM mixing in neutrino sector. The subleading corrections to charged fermions are
negligible with respect to the LO predictions. The final results are not altered in order of
magnitude even in the expressions. In the end we show the phenomenological numerical results predicted by the model. Future long base line neutrino
experiments with higher precision is possible to verify or falsify the result about leptonic CP violating phase predicted in the model.
The neutrinoless double beta decay experiment is also a tool for testing the model, which can discriminate between the NH spectrum and the IH one.

\section*{Acknowledgements}
One of the authors (Y. Z) would like to thank Y.L. Zhou for useful discussions. This work is partially supported by the Foundation of National Key
Program for Basic Research of China (2001CCB01000).
\appendix
\begin{appendix}
\renewcommand{\theequation}{{A}\thesection\arabic{equation}}
\setcounter{equation}{0}
\section*{APPENDIX A: $S_{4}$ GROUP AND REPRESENTATIONS}\label{A}
The discrete flavor group $S_{4}$, permutation group of four objects, has 24 elements. The two generators \textit{S} and \textit{T} in different
irreducible presentations are given as follows
\begin{eqnarray}
  & & 1_{1} : S = 1,  \quad  T=1 \\
  & & 1_{2} : S = -1, \quad  T=1 \\
  & & 2  :
S = \left(
  \begin{array}{cc}
  0 & 1 \\
  1 & 0 \\
  \end{array}
  \right),\quad
T=\left(
  \begin{array}{cc}
  \omega & 0 \\
  0 & \omega^{2} \\
  \end{array}
  \right)\\
& & 3_{1} :
S = \frac{1}{3}\left(
  \begin{array}{ccc}
  -1 & 2\omega & 2\omega^{2} \\
  2\omega & 2\omega^{2} & -1 \\
  2\omega^{2} & -1 & 2\omega \\
  \end{array}
  \right),\quad
T = \left(
  \begin{array}{ccc}
  1 & 0 & 0 \\
  0 & \omega^{2} & 0 \\
  0 & 0 & \omega \\
  \end{array}
  \right)\\
& & 3_{2} :
S = \frac{1}{3}\left(
  \begin{array}{ccc}
  1 & -2\omega & -2\omega^{2} \\
  -2\omega & -2\omega^{2} & 1 \\
  -2\omega^{2} & 1 & -2\omega \\
  \end{array}
  \right),\quad
T = \left(
  \begin{array}{ccc}
  1 & 0 & 0 \\
  0 & \omega^{2} & 0 \\
  0 & 0 & \omega \\
  \end{array}
  \right)
\end{eqnarray}
where $\omega=e^{2\pi i/3}=(i\sqrt{3}-1)/2$. The five conjugate classes of $S_{4}$ group can be written as
\begin{eqnarray}
  & & \mathcal {C}_{1} : 1 \nonumber\\
  & & \mathcal {C}_{2} : S^{2}, \quad TS^{2}T^{2}, \quad T^{2}S^{2}T \nonumber\\
  & & \mathcal {C}_{3} : T, \quad T^{2}, \quad S^{2}T, \quad S^{2}T^{2}, \quad STST^{2}, \quad STS, \quad TS^{2}, \quad T^{2}S^{2}  \nonumber\\
  & & \mathcal {C}_{4} : ST^{2}, \quad T^{2}S,\quad TST, \quad TSTS^{2}, \quad STS^{2}, \quad S^{2}TS \nonumber\\
  & & \mathcal {C}_{5} : S, \quad TST^{2}, \quad ST, \quad TS, \quad S^{3}, \quad S^{3}T^{2}
\end{eqnarray}

\begin{table}[!htbp]
  \centering
  \caption{The character table of $S_{4}$ group}\label{tab3}
\begin{tabular}{cccccc}
  \hline
  $ $ & $1_{1}$ & $1_{2}$ & 2 & $3_{1}$ & $3_{2}$ \\
  \hline
  $\mathcal{C}_{1}$ & 1 & 1 & 2 & 3 & 3 \\
  $\mathcal{C}_{2}$ & 1 & 1 & 2 & $-1$ & $-1$ \\
  $\mathcal{C}_{3}$ & 1 & 1 & $-1$ & 0 & 0 \\
  $\mathcal{C}_{4}$ & 1 & $-1$ & 0 & 1 & $-1$ \\
  $\mathcal{C}_{5}$ & 1 & $-1$ & 0 & $-1$ & 1 \\
  \hline
\end{tabular}
\end{table}
The character table of $S_{4}$ group are shown in \hyperref[tab3]{Table 3} and the multiplication rules between various irreducible representations
are shown in Eq.~\eqref{eq:rule}. Taking into account of the generators \textit{S} and \textit{T} in different representations, one may obtain the
representation matrices of all elements. The explicit expressions of $S_{4}$ elements in different representations can be found in
Refs.~\cite{Melo/2010,Ding/2010S4}, especially the subgroups of $S_{4}$ are thorough classified in~\cite{Ding/2010S4}. In this basis we
can straightforwardly obtain the decomposition of the product representations and the Clebsch-Gordan factors. To be specific the product rules
of $S_{4}$ group, with $\psi_{i}$, $\varphi_{i}$ as the elements of the first and second representation of the product, respectively, are given
as follows
\begin{eqnarray}
& & \star \quad 1_{1} \otimes r = r \otimes 1_{1} = r \qquad \text{with $r$ being any representation} \\
& & \star \quad 1_{2} \otimes 1_{2} = 1_{1} \sim \psi\varphi \\
& & \star \quad 1_{2} \otimes 2 = 2  \sim \left(
    \begin{array}{c}
    \psi\varphi_{1} \\
   -\psi\varphi_{2} \\
    \end{array}
    \right) \\
& & \star \quad 1_{2}\otimes 3_{1} = 3_{2} \sim \left(
    \begin{array}{c}
    \psi\varphi_{1} \\
    \psi\varphi_{2} \\
    \psi\varphi_{3} \\
    \end{array}
    \right) \\
& & \star \quad 1_{2}\otimes 3_{2} = 3_{1} \sim \left(
    \begin{array}{c}
    \psi\varphi_{1} \\
    \psi\varphi_{2} \\
    \psi\varphi_{3} \\
    \end{array}
    \right)
\end{eqnarray}
The product rules with two-dimensional representation are as follows:
\begin{eqnarray}
& & \star \quad 2 \otimes 2 = 1_{1} \oplus 1_{2} \oplus 2          \nonumber\\
& & \quad \quad 1_{1} \sim \psi_{1}\varphi_{2}+\psi_{2}\varphi_{1} ,
\quad \quad 1_{2} \sim \psi_{1}\varphi_{2}-\psi_{2}\varphi_{1} \nonumber\\
& & \quad \quad 2 \sim \left(
    \begin{array}{c}
    \psi_{2}\varphi_{2} \\
    \psi_{1}\varphi_{1} \\
    \end{array}
    \right)  \\
& & \star \quad 2 \otimes 3_{1} = 3_{1} \oplus 3_{2}  \nonumber\\
& & \quad \quad 3_{1} \sim \left(
\begin{array}{c}
    \psi_{1}\varphi_{2}+\psi_{2}\varphi_{3} \\
    \psi_{1}\varphi_{3}+\psi_{2}\varphi_{1} \\
    \psi_{1}\varphi_{1}+\psi_{2}\varphi_{2} \\
    \end{array}
    \right) ,
\quad \quad 3_{2} \sim \left(
    \begin{array}{c}
    \psi_{1}\varphi_{2}-\psi_{2}\varphi_{3} \\
    \psi_{1}\varphi_{3}-\psi_{2}\varphi_{1} \\
    \psi_{1}\varphi_{1}-\psi_{2}\varphi_{2} \\
    \end{array}
    \right) \\
& & \star \quad 2 \otimes 3_{2} = 3_{1} \oplus 3_{2}  \nonumber\\
& & \quad \quad 3_{1} \sim \left(
\begin{array}{c}
    \psi_{1}\varphi_{2}-\psi_{2}\varphi_{3} \\
    \psi_{1}\varphi_{3}-\psi_{2}\varphi_{1} \\
    \psi_{1}\varphi_{1}-\psi_{2}\varphi_{2} \\
    \end{array}
    \right) ,
\quad \quad 3_{2} \sim \left(
    \begin{array}{c}
    \psi_{1}\varphi_{2}+\psi_{2}\varphi_{3} \\
    \psi_{1}\varphi_{3}+\psi_{2}\varphi_{1} \\
    \psi_{1}\varphi_{1}+\psi_{2}\varphi_{2} \\
    \end{array}
    \right)
\end{eqnarray}
and the product rules with three-dimensional representation are as follows
\begin{eqnarray}
& & \star \quad 3_{1}\otimes 3_{1} = 3_{2} \otimes 3_{2}=1_{1} \oplus 2 \oplus 3_{1} \oplus 3_{2} \nonumber\\
& & \quad \quad 1_{1} \sim \psi_{1}\varphi_{1}+\psi_{2}\varphi_{3}+\psi_{3}\varphi_{2} \nonumber\\
& & \quad \quad 2 \sim \left(
  \begin{array}{c}
  \psi_{2}\varphi_{2}+\psi_{3}\varphi_{1}+\psi_{1}\varphi_{3} \\
  \psi_{3}\varphi_{3}+\psi_{1}\varphi_{2}+\psi_{2}\varphi_{1} \\
  \end{array}
  \right)\nonumber\\
& & \quad \quad 3_{1} \sim \left(
  \begin{array}{c}
  2\psi_{1}\varphi_{1}-\psi_{2}\varphi_{3}-\psi_{3}\varphi_{2} \\
  2\psi_{3}\varphi_{3}-\psi_{1}\varphi_{2}-\psi_{2}\varphi_{1} \\
  2\psi_{2}\varphi_{2}-\psi_{3}\varphi_{1}-\psi_{1}\varphi_{3} \\
  \end{array}
  \right),
\quad \quad  3_{2} \sim \left(
  \begin{array}{c}
  \psi_{2}\varphi_{3}-\psi_{3}\varphi_{2} \\
  \psi_{1}\varphi_{2}-\psi_{2}\varphi_{1} \\
  \psi_{3}\varphi_{1}-\psi_{1}\varphi_{3} \\
  \end{array}
  \right)
\end{eqnarray}
\begin{eqnarray}
& & \star \quad 3_{1}\otimes 3_{2} = 1_{2} \oplus 2 \oplus 3_{1} \oplus 3_{2} \nonumber\\
& & \quad \quad 1_{2} \sim \psi_{1}\varphi_{1}+\psi_{2}\varphi_{3}+\psi_{3}\varphi_{2} \nonumber\\
& & \quad \quad 2 \sim \left(
  \begin{array}{c}
  \psi_{2}\varphi_{2}+\psi_{3}\varphi_{1}+\psi_{1}\varphi_{3} \\
  -\psi_{3}\varphi_{3}-\psi_{1}\varphi_{2}-\psi_{2}\varphi_{1} \\
  \end{array}
  \right)   \nonumber\\
& & \quad \quad 3_{1}\sim
    \left(
  \begin{array}{c}
  \psi_{2}\varphi_{3}-\psi_{3}\varphi_{2} \\
  \psi_{1}\varphi_{2}-\psi_{2}\varphi_{1} \\
  \psi_{3}\varphi_{1}-\psi_{1}\varphi_{3} \\
  \end{array}
  \right),
\quad \quad  3_{2} \sim\left(
  \begin{array}{c}
  2\psi_{1}\varphi_{1}-\psi_{2}\varphi_{3}-\psi_{3}\varphi_{2} \\
  2\psi_{3}\varphi_{3}-\psi_{1}\varphi_{2}-\psi_{2}\varphi_{1} \\
  2\psi_{2}\varphi_{2}-\psi_{3}\varphi_{1}-\psi_{1}\varphi_{3} \\
  \end{array}
  \right)
\end{eqnarray}
\end{appendix}
\begin{appendix}
\renewcommand{\theequation}{{B}\thesection\arabic{equation}}
\setcounter{equation}{0}
\section*{APPENDIX B: Corrections to vacuum alignment}\label{B}
The subleading corrections to LO vacuum alignment stem from higher dimensional operators which are suppressed by one or more power
of $1/\Lambda$. The modified driving superpotential will consist of leading order term $w^{0}_{d}$, which is just Eq.~\eqref{eq:LO wd}, and
correctional term $\delta w_{d}$, which comes from all invariant operators linear in the driving fields that suppressed by $1/\Lambda$
at least one power
\begin{equation}\label{eq:Wd AppB}
  \qquad w_{d}=w^{0}_{d}+\delta w_{d}
\end{equation}
The correctional term $\delta w_{d}$ contains all contractional invariant subdominant operators under the flavor symmetry
$S_{4}\times Z_{4}\times Z_{6}\times Z_{5}\times Z_{2}$
\begin{eqnarray}\label{dwdB}
& & \delta w_{d}=\sum_{i=1}^{5}\frac{a_{i}}{\Lambda}\mathcal {O}^{\varphi_{0}}_{i}
  +\sum_{i=1}^{8}\frac{b_{i}}{\Lambda}\mathcal {O}^{\chi_{0}}_{i}+\frac{c}{\Lambda}\mathcal {O}^{\eta_{0}}
  +\sum_{i=1}^{3}\frac{m_{i}}{\Lambda}\mathcal {O}^{\sigma_{0}}
  +\frac{k}{\Lambda}\mathcal {O}^{\xi_{0}}\nonumber\\
& &\qquad\quad  +\sum_{i=1}^{12}\frac{s_{i}}{\Lambda^{2}}\mathcal {O}^{\phi_{0}}_{i}
  +\sum_{i=1}^{12}\frac{t_{i}}{\Lambda^{2}}\mathcal {O}^{\Delta_{0}}_{i}
  +\sum_{i=1}^{11}\frac{u_{i}}{\Lambda^{2}}\mathcal {O}^{\zeta_{0}}_{i}
\end{eqnarray}
in which the complex coefficients $a_{i}$, $b_{i}$, $c$, $d_{i}$, $m$, $k$, $s_{i}$, $t_{i}$ and $u_{i}$ are all of order one but cannot be specific
determined according to the flavor symmetry. Operators $\mathcal {O}_{i}$ represent all the subdominant invariant contractional operators under the
symmetry group $S_{4}\times Z_{4} \times Z_{6}\times Z_{5}\times Z_{2}$
\begin{eqnarray}
& & \mathcal{O}^{\varphi_{0}}_{1} = (\varphi_{0}\chi)_{2}(\phi\phi)_{2}                  , \quad
    \mathcal{O}^{\varphi_{0}}_{2} = (\varphi_{0}\chi)_{3_{1}}(\phi\phi)_{3_{1}}          , \quad
    \mathcal{O}^{\varphi_{0}}_{3} = (\varphi_{0}\chi)_{3_{1}}(\phi\Delta)_{3_{1}}         \nonumber\\
& & \mathcal{O}^{\varphi_{0}}_{4} = (\varphi_{0}\chi)_{3_{2}}(\phi\Delta)_{3_{2}}        , \quad
    \mathcal{O}^{\varphi_{0}}_{5} = (\varphi_{0}\chi)_{2}(\Delta\Delta)_{2}              , \quad
\end{eqnarray}
\begin{eqnarray}
& & \mathcal{O}^{\chi_{0}}_{1} = (\chi_{0}\chi)_{1_{1}}(\phi\phi)_{1_{1}}                       , \quad
    \mathcal{O}^{\chi_{0}}_{2} = (\chi_{0}\chi)_{2}(\phi\phi)_{2}                               , \quad
    \mathcal{O}^{\chi_{0}}_{3} = (\chi_{0}\chi)_{3_{1}}(\phi\phi)_{3_{1}}                         \nonumber\\
& & \mathcal{O}^{\chi_{0}}_{4} = (\chi_{0}\chi)_{3_{1}}(\phi\Delta)_{3_{1}}                     , \quad
    \mathcal{O}^{\chi_{0}}_{5} = (\chi_{0}\chi)_{3_{2}}(\phi\Delta)_{3_{2}}                       \nonumber\\
& & \mathcal{O}^{\chi_{0}}_{6} = (\chi_{0}\chi)_{1_{1}}(\Delta\Delta)_{1_{1}}                   , \quad
    \mathcal{O}^{\chi_{0}}_{7} = (\chi_{0}\chi)_{2}(\Delta\Delta)_{2}                           , \quad
    \mathcal{O}^{\chi_{0}}_{8} = (\chi_{0}\chi)_{1_{1}}\zeta\zeta
\end{eqnarray}
\begin{eqnarray}
  & & \mathcal{O}^{\eta_{0}} = \eta_{0}\chi(\phi\Delta)_{3_{2}}
\end{eqnarray}
\begin{eqnarray}
& & \mathcal{O}^{\sigma_{0}}_{1} = \sigma_{0}(\varphi\chi)_{3_{1}}\phi,\quad
    \mathcal{O}^{\sigma_{0}}_{2} = \sigma_{0}(\varphi\chi)_{2}\Delta,\quad
    \mathcal{O}^{\sigma_{0}}_{3} = \sigma_{0}(\eta\chi)_{3_{1}}\phi
\end{eqnarray}
\begin{eqnarray}
& & \mathcal{O}^{\xi_{0}} =  \xi_{0}(\xi\chi)_{3_{1}}\vartheta, 
\end{eqnarray}
\begin{eqnarray}
& & \mathcal{O}^{\phi_{0}}_{1} = \phi_{0}(\phi\Delta)_{3_{2}}\zeta\zeta                                                               , \quad
    \mathcal{O}^{\phi_{0}}_{2} = \phi_{0}\big((\phi\phi)_{2}(\phi\phi)_{3_{1}}\big)_{3_{2}}                                            , \quad
    \mathcal{O}^{\phi_{0}}_{3} = \phi_{0}\big((\phi\phi)_{3_{1}}(\phi\phi)_{3_{1}}\big)_{3_{2}}                                        , \nonumber\\
& & \mathcal{O}^{\phi_{0}}_{4} = \phi_{0}(\phi\phi)_{1_{1}}(\phi\Delta)_{3_{2}}                                                        , \quad
    \mathcal{O}^{\phi_{0}}_{5} = \phi_{0}\big((\phi\phi)_{2}(\phi\Delta)_{3_{1}}\big)_{3_{2}}                                          , \quad
    \mathcal{O}^{\phi_{0}}_{6} = \phi_{0}\big((\phi\phi)_{2}(\phi\Delta)_{3_{2}}\big)_{3_{2}}                                          , \nonumber\\
& & \mathcal{O}^{\phi_{0}}_{7} = \phi_{0}\big((\phi\phi)_{3_{1}}(\phi\Delta)_{3_{1}}\big)_{3_{2}}                                      , \quad
    \mathcal{O}^{\phi_{0}}_{8} = \phi_{0}\big((\phi\phi)_{3_{1}}(\phi\Delta)_{3_{2}}\big)_{3_{2}}                                      , \quad
    \mathcal{O}^{\phi_{0}}_{9} = \phi_{0}\big((\phi\phi)_{3_{1}}(\Delta\Delta)_{2}\big)_{3_{2}}                                        , \nonumber\\
& & \mathcal{O}^{\phi_{0}}_{10} = \phi_{0}(\phi\Delta)_{3_{2}}(\Delta\Delta)_{1_{1}}                                                   , \quad
    \mathcal{O}^{\phi_{0}}_{11} = \phi_{0}\big((\phi\Delta)_{3_{1}}(\Delta\Delta)_{2}\big)_{3_{2}}                                     , \quad
    \mathcal{O}^{\phi_{0}}_{12} = \phi_{0}\big((\phi\Delta)_{3_{2}}(\Delta\Delta)_{2}\big)_{3_{2}}
\end{eqnarray}
\begin{eqnarray}
& & \mathcal{O}^{\Delta_{0}}_{1} = \Delta_{0}(\phi\phi)_{1_{1}}(\phi\phi)_{2}                        ,  \quad
    \mathcal{O}^{\Delta_{0}}_{2} = \Delta_{0} \big((\phi\phi)_{2}(\phi\phi)_{2} \big)_{2}            ,  \quad
    \mathcal{O}^{\Delta_{0}}_{3} = \Delta_{0} \big((\phi\phi)_{3_{1}}(\phi\phi)_{3_{1}} \big)_{2}    ,  \nonumber\\
& & \mathcal{O}^{\Delta_{0}}_{4} = \Delta_{0} \big((\phi\phi)_{3_{1}}(\phi\Delta)_{3_{1}} \big)_{2}  ,  \quad
    \mathcal{O}^{\Delta_{0}}_{5} = \Delta_{0} \big((\phi\phi)_{3_{1}}(\phi\Delta)_{3_{2}} \big)_{2}  ,  \quad
    \mathcal{O}^{\Delta_{0}}_{6} = \Delta_{0}(\phi\phi)_{1_{1}}(\Delta\Delta)_{2}                    ,  \nonumber\\
& & \mathcal{O}^{\Delta_{0}}_{7} = \Delta_{0}(\phi\phi)_{2}(\Delta\Delta)_{1_{1}}                    ,  \quad
    \mathcal{O}^{\Delta_{0}}_{8} = \Delta_{0}\big((\phi\phi)_{2}(\Delta\Delta)_{2}\big)_{2}          ,  \quad
    \mathcal{O}^{\Delta_{0}}_{9} = \Delta_{0}(\phi\phi)_{2}\zeta\zeta                               ,  \nonumber\\
& & \mathcal{O}^{\Delta_{0}}_{10} = \Delta_{0}(\Delta\Delta)_{1_{1}}(\Delta\Delta)_{2}               ,  \quad
    \mathcal{O}^{\Delta_{0}}_{11} = \Delta_{0}(\Delta\Delta)_{2}(\Delta\Delta)_{2}                   ,  \quad
    \mathcal{O}^{\Delta_{0}}_{12} = \Delta_{0}(\Delta\Delta)_{2}\zeta\zeta                          ,  \quad
\end{eqnarray}
\begin{eqnarray}
& & \mathcal{O}^{\zeta_{0}}_{1} = \zeta_{0}\zeta\zeta(\phi\phi)_{1_{1}}             ,\quad
    \mathcal{O}^{\zeta_{0}}_{2} = \zeta_{0}\zeta\zeta(\Delta\Delta)_{1_{1}}         ,\quad
    \mathcal{O}^{\zeta_{0}}_{3} = \zeta_{0}\zeta\zeta\zeta\zeta                    ,\nonumber\\
& & \mathcal{O}^{\zeta_{0}}_{4} =  \zeta_{0}(\phi\phi)_{1_{1}}(\phi\phi)_{1_{1}}                     ,\quad
    \mathcal{O}^{\zeta_{0}}_{5} =  \zeta_{0}\big((\phi\phi)_{2}(\phi\phi)_{2}\big)_{1_{1}}           ,\quad
    \mathcal{O}^{\zeta_{0}}_{6} =  \zeta_{0}\big((\phi\phi)_{3_{1}}(\phi\phi)_{3_{1}}\big)_{1_{1}}   ,\nonumber\\
& & \mathcal{O}^{\zeta_{0}}_{7} =  \zeta_{0}\big((\phi\phi)_{3_{1}}(\phi\Delta)_{3_{1}}\big)_{1_{1}} ,\quad
    \mathcal{O}^{\zeta_{0}}_{8} =  \zeta_{0}(\phi\phi)_{1_{1}}(\Delta\Delta)_{1_{1}}                 ,\quad
    \mathcal{O}^{\zeta_{0}}_{9} =  \zeta_{0}\big((\phi\phi)_{2}(\Delta\Delta)_{2}\big)_{1_{1}}           ,\nonumber\\
& & \mathcal{O}^{\zeta_{0}}_{10} =  \zeta_{0}(\Delta\Delta)_{1_{1}}(\Delta\Delta)_{1_{1}}             ,\quad
    \mathcal{O}^{\zeta_{0}}_{11} =  \zeta_{0}\big((\Delta\Delta)_{2}(\Delta\Delta)_{2}\big)_{1_{1}}
\end{eqnarray}
The sub-dominate term $\delta w_{d}$ induces shifted VEVs of all flavon fields, and we can rewrite the modified vacuum alignment as following
\begin{eqnarray}
  & & \langle \varphi \rangle = \left(
      \begin{array}{c}
      \delta v_{\varphi_{1}} \\
      v_{\varphi}+\delta v_{\varphi_{2}} \\
      \delta v_{\varphi_{3}} \\
      \end{array}
      \right)     , \quad
      \langle \eta \rangle = \left(
      \begin{array}{c}
      \delta v_{\eta_{1}} \\
      v_{\eta} \\
      \end{array}
      \right)    , \quad
      \langle \chi \rangle = \left(
      \begin{array}{c}
      \delta v_{\chi_{1}} \\
      \delta v_{\chi_{2}} \\
      v_{\chi}+\delta v_{\chi_{3}} \\
      \end{array}
      \right), \quad \langle \xi \rangle = \left(
      \begin{array}{c}
      v_{\xi}+\delta v_{\xi_{1}} \\
      \delta v_{\xi_{2}} \\
      \delta v_{\xi_{3}} \\
      \end{array}
      \right),\nonumber\\
& &    \langle \rho \rangle = v_{\rho} ,\quad
      \langle \sigma \rangle = v_{\sigma}+\delta v_{\sigma}, \quad
      \langle \vartheta \rangle = v_{\vartheta} ,
\end{eqnarray}
\begin{eqnarray}
  & & \langle \phi \rangle = \left(
      \begin{array}{c}
      v_{\phi}+\delta v_{\phi_{1}} \\
      v_{\phi}+\delta v_{\phi_{2}} \\
      v_{\phi}+\delta v_{\phi_{3}} \\
      \end{array}
      \right), \quad
      \langle \Delta \rangle = \left(
      \begin{array}{c}
      v_{\Delta}+\delta v_{\Delta_{1}} \\
      v_{\Delta}+\delta v_{\Delta_{2}} \\
      \end{array}
      \right)       , \quad
      \langle \zeta \rangle = v_{\zeta},
\end{eqnarray}
where the shifts $\delta v_{\eta_{2}}$, $\delta v_{\vartheta}$, $\delta v_{\rho}$ and $\delta v_{\zeta}$ have been absorbed
into the redefinition of the undetermined $v_{\eta}$, $v_{\vartheta}$, $v_{\rho}$ and $v_{\zeta}$ respectively. With only
terms linear in the shift $\delta v$ retained and ignoring the $\delta v /\Lambda$ terms, the new minimization equations are still derived by the zeros
of F-terms, i.e. the first derivative of new $w_{d}$ in Eq.~\eqref{eq:Wd AppB} with respect to all driving fields. First the minimization equations for
the set $\Phi^{e}$ are showed as follows
\begin{eqnarray}\label{eq:ve1 AppB}
  & & -2g_{1}v_{\varphi}\delta v_{\varphi_{3}}+g_{2}(v_{\varphi}\delta v_{\eta_{1}}
      +v_{\eta}\delta v_{\varphi_{3}})+\frac{v_{\chi}v^{2}_{\phi}}{\Lambda}A_{1} = 0                           \nonumber\\
  & & (4g_{1}v_{\varphi}+g_{2}v_{\eta})\delta v_{\varphi_{2}}+\frac{v_{\chi}v^{2}_{\phi}}{\Lambda}A_{2} = 0    \nonumber\\
  & & (-2g_{1}v_{\varphi}+g_{2}v_{\eta})\delta v_{\varphi_{1}}+\frac{v_{\chi}v^{2}_{\phi}}{\Lambda}A_{3} = 0   \nonumber\\
  & & g_{3}M_{\chi}\delta v_{\chi_{1}}+g_{4}(v_{\eta}\delta v_{\varphi_{3}}-v_{\varphi}\delta v_{\eta_{1}})
      +\frac{v_{\chi}v^{2}_{\phi}}{\Lambda}B_{1} = 0                                                                     \nonumber\\
  & & g_{3}M_{\chi}\delta v_{\chi_{3}}+g_{4}v_{\eta}\delta v_{\varphi_{2}}+\frac{v_{\chi}v^{2}_{\phi}}{\Lambda}B_{2} = 0 \nonumber\\
  & & g_{3}M_{\chi}\delta v_{\chi_{2}}+g_{4}v_{\eta}\delta v_{\varphi_{1}}+\frac{v_{\chi}v^{2}_{\phi}}{\Lambda}B_{3} = 0 \nonumber\\
  & & 2h_{1}v_{\varphi}\delta v_{\varphi_{3}}+2h_{2}v_{\eta}\delta v_{\eta_{1}}+\frac{v_{\chi}v_{\phi}v_{\Delta}}{\Lambda}C = 0
\end{eqnarray}
where the coefficients $A_{1,2,3}$, $B_{1,2,3}$ and C stand for the linear combinations of sub-leading coefficients
\begin{eqnarray}
  & &  A_{1} =  3a_{1}-2a_{3}\frac{v_{\Delta}}{v_{\phi}}+a_{5}\frac{v^{2}_{\Delta}}{v^{2}_{\phi}},
  \quad  A_{2} =  2a_{3}\frac{v_{\Delta}}{v_{\phi}},
  \quad  A_{3} =  -3a_{1}-a_{5}\frac{v^{2}_{\Delta}}{v^{2}_{\phi}}                  \nonumber\\
  & &  B_{1} =  3b_{2}-2b_{4}\frac{v_{\Delta}}{v_{\phi}}+b_{6}\frac{v_{\zeta}}{v_{\phi}}+b_{8}\frac{v^{2}_{\Delta}}{v^{2}_{\phi}},
  \quad  B_{2} =  3b_{1}-2b_{4}\frac{v_{\Delta}}{v_{\phi}}+b_{6}\frac{v_{\zeta}}{v_{\phi}}+2b_{7}\frac{v^{2}_{\Delta}}{v^{2}_{\phi}} \nonumber\\
  & &  B_{3} =  3b_{2}+4b_{4}\frac{v_{\Delta}}{v_{\phi}}+b_{8}\frac{v^{2}_{\Delta}}{v^{2}_{\phi}},
  \quad C = 0
\end{eqnarray}

The solutions for Eq.~\eqref{eq:ve1 AppB} can be easily obtained as follows
\begin{eqnarray}\label{eq:delta Ve1}
  & & \delta v_{\varphi_{1}} = \frac{A_{3}}{4g_{1}}\frac{v_{\chi}}{\Lambda}\frac{v^{2}_{\phi}}{v_{\varphi}},
  \quad \delta v_{\varphi_{2}} = -\frac{A_{2}}{2g_{1}}\frac{v_{\chi}}{\Lambda}\frac{v^{2}_{\phi}}{v_{\varphi}},
  \quad \delta v_{\varphi_{3}} = \frac{2g_{1}h_{2}A_{1}}{8g^{2}_{1}h_{2}-g^{2}_{2}h_{1}}
  \frac{v_{\chi}}{\Lambda}\frac{v^{2}_{\phi}}{v_{\varphi}}                                          \nonumber\\
  & & \delta v_{\chi_{1}} = -\Big[\frac{B_{1}}{g_{3}}+\frac{A_{1}g_{4}}{g^{2}_{2}h_{1}-8g^{2}_{1}h_{2}}
  \Big(\frac{4g^{2}_{1}h_{2}}{g_{2}g_{3}}+\frac{g_{2}h_{1}}{g_{3}}\Big)\Big]\frac{v_{\chi}}{\Lambda}\frac{v^{2}_{\phi}}{M_{\chi}}\\
  & & \delta v_{\chi_{2}} = \Big(\frac{g_{4}A_{3}}{2g_{2}g_{3}}-\frac{B_{3}}{g_{3}}\Big)\frac{v_{\chi}}{\Lambda}\frac{v^{2}_{\phi}}{M_{\chi}},
  \quad \delta v_{\chi_{3}} = \Big(-\frac{g_{4}A_{2}}{g_{3}g_{2}}-\frac{B_{2}}{g_{3}}\Big)\frac{v_{\chi}}{\Lambda}\frac{v^{2}_{\phi}}{M_{\chi}},
  \quad \delta v_{\eta_{1}} = \frac{2g_{1}h_{1}A_{1}}{g^{2}_{2}h_{1}-8g^{2}_{1}h_{2}}\frac{v_{\chi}}{\Lambda}\frac{v^{2}_{\phi}}{v_{\eta}}   \nonumber
\end{eqnarray}

The minimization equation for the shifted VEV $\delta v_{\sigma}$ is obtained in the same way as follows
\begin{eqnarray}\label{eq:ve21 AppB}
& & 2q_{1}v_{\sigma}\delta v_{\sigma}+ \frac{v_{\varphi}v_{\chi}v_{\phi}}{\Lambda}M=0
\end{eqnarray}
where the coefficients M is naively as
\begin{eqnarray}
&¡¡& M=m_{1}-m_{2}\frac{v_{\eta}}{v_{\varphi}}
\end{eqnarray}
Then Eq.~\eqref{eq:ve21 AppB} admits the following solutions
\begin{eqnarray}\label{eq:delta Ve21}
& & \delta v_{\sigma}=-\frac{M}{2q_{1}}\frac{v_{\phi}}{\Lambda}\frac{v_{\varphi}v_{\chi}}{v_{\sigma}}
\end{eqnarray}

The equations for corrections $\delta v_{\xi_{1,2,3}}$
are simple as follows
\begin{eqnarray}\label{eq:ve22 AppB}
  & &  4r_{1}v_{\xi}\delta v_{\xi_{1}}+r_{2}v_{\rho}\delta v_{\xi_{1}} = 0            \nonumber\\
  & & -2r_{1}v_{\xi}\delta v_{\xi_{3}}+r_{2}v_{\rho}\delta v_{\xi_{3}} = 0           \nonumber\\
  & & -2r_{1}v_{\xi}\delta v_{\xi_{2}}+r_{2}v_{\rho}\delta v_{\xi_{2}}+K\frac{v_{\xi}v_{\varphi}v_{\vartheta}}{\Lambda} = 0
\end{eqnarray}
where the coefficient K is easily solved as
\begin{equation}\label{eq:K}
  \qquad K=k_{1}+k_{2}\frac{v_{\eta}}{v_{\varphi}}
\end{equation}
and the solutions to the above minimization equations~\eqref{eq:ve22 AppB} are
\begin{eqnarray}\label{eq:delta Ve22}
  & & \delta v_{\xi_{1}}= 0,
  \quad \delta v_{\xi_{2}}= -\frac{K}{4r_{1}}\frac{v_{\varphi}v_{\vartheta}}{\Lambda},
  \quad \delta v_{\xi_{3}}= 0
\end{eqnarray}
Despite of some shifts, $\delta v_{\xi_{1,3}}$, are exactly zero, most of the shifted VEVs $\delta v_{\Phi^{e}}$ are all of order in the
interval [$\lambda^{6}_{c}\Lambda$, $\lambda^{4}_{c}\Lambda$] from Eqs.~\eqref{eq:delta Ve1},~\eqref{eq:delta Ve21} and~\eqref{eq:delta Ve22},
which imply the relative order of shifted VEVs with respect the LO VEVs of $\Phi^{e}$ are in the interval [$\lambda^{4}_{c}$, $\lambda^{2}_{c}$], i.e.,
$\delta v_{\Phi^{e}}/v_{\Phi^{e}}\in [\lambda^{4}_{c}, \lambda^{2}_{c}]$. As illuminated in the end of Section \ref{S3}, the subleading
corrections to $\langle\Phi^{e}\rangle$ should be smaller than the mass ratio $\frac{m_{\mu}}{m_{\tau}}$ or, more strictly $\frac{m_{e}}{m_{\tau}}$.
The results above have shown the corrections are suitable for the model, and LO VEVs in
Eqs.~\eqref{eq:Ve11}~\eqref{eq:Ve12} and~\eqref{eq:Ve2} are stable solutions even under NLO corrections.

At last for the $\Phi^{\nu}$ sector we have the minimization equations
\begin{eqnarray}\label{eq:del}
& & f_{1}[v_{\phi}(\delta v_{\Delta_{1}}-\delta v_{\Delta_{2}})+v_{\Delta}(\delta v_{\phi_{2}}-\delta v_{\phi_{3}})]
         +\frac{v^{4}_{\phi}}{\Lambda^{2}}C_{1} = 0                                          \nonumber\\
& & f_{1}[v_{\phi}(\delta v_{\Delta_{1}}-\delta v_{\Delta_{2}})+v_{\Delta}(\delta v_{\phi_{1}}-\delta v_{\phi_{2}})]
         +\frac{v^{4}_{\phi}}{\Lambda^{2}}C_{1} = 0                                          \nonumber\\
& & f_{1}[v_{\phi}(\delta v_{\Delta_{1}}-\delta v_{\Delta_{2}})+v_{\Delta}(\delta v_{\phi_{3}}-\delta v_{\phi_{1}})]
         +\frac{v^{4}_{\phi}}{\Lambda^{2}}C_{1} = 0                                          \nonumber\\
& & 2f_{2}v_{\phi}(\delta v_{\phi_{1}}+\delta v_{\phi_{2}}+\delta v_{\phi_{3}})+2f_{3}v_{\Delta}\delta v_{\Delta_{1}}
   +\frac{v^{4}_{\phi}}{\Lambda^{2}}C_{2}=0                                                            \nonumber\\
& & 2f_{2}v_{\phi}(\delta v_{\phi_{1}}+\delta v_{\phi_{2}}+\delta v_{\phi_{3}})+2f_{3}v_{\Delta}\delta v_{\Delta_{2}}
   +\frac{v^{4}_{\phi}}{\Lambda^{2}}C_{2}=0                                                            \nonumber\\
& & 2f_{5}v_{\phi}(\delta v_{\phi_{1}}+\delta v_{\phi_{2}}+\delta v_{\phi_{3}})+2f_{6}v_{\Delta}(\delta v_{\Delta_{1}}+\delta v_{\Delta_{2}})
   +\frac{v^{2}_{\phi}v_{\zeta}^{2}}{\Lambda^{2}}C_{3}=0
\end{eqnarray}
where the coefficients $C_{1}$, $C_{2}$ and $C_{3}$ are
\begin{eqnarray}\label{eq:C}
& & C_{1} = 0                              \nonumber\\
& & C_{2} = 9(t_{1}+t_{2})+3(t_{6}+2t_{7}+t_{8})\frac{v^{2}_{\Delta}}{v^{2}_{\phi}}+3t_{9}\frac{v^{2}_{\zeta}}{v^{2}_{\phi}}
            +(2t_{10}+t_{11})\frac{v^{4}_{\Delta}}{v^{4}_{\phi}}+t_{12}\frac{v^{2}_{\Delta}v^{2}_{\zeta}}{v^{4}_{\phi}}      \nonumber\\
& & C_{3}  = 3u_{1}+2u_{2}\frac{v^{2}_{\Delta}}{v^{2}_{\phi}}+u_{3}\frac{v_{\zeta}^{2}}{v^{2}_{\phi}}
            +9(u_{4}+2u_{5})\frac{v^{2}_{\phi}}{v^{2}_{\zeta}}+6(u_{8}+u_{9})\frac{v^{2}_{\Delta}}{v^{2}_{\zeta}}
            +2(2u_{10}+u_{11})\frac{v^{4}_{\Delta}}{v^{2}_{\phi}v^{2}_{\zeta}}
\end{eqnarray}
The solutions to the Eqs.~\eqref{eq:del} are also easily obtained as follows
\begin{eqnarray}\label{eq:delta Vf1}
& & \delta v_{\phi_{1}} = \delta v_{\phi_{2}} = \delta v_{\phi_{3}} =
\frac{2C_{2}f_{6}v^{2}_{\phi}-C_{3}f_{3}v^{2}_{\zeta}}{(6f_{3}f_{5}-12f_{2}f_{6})v_{\phi}}\frac{v^{2}_{\phi}}{\Lambda^{2}}         \nonumber\\
& & \delta v_{\Delta_{1}} = \delta v_{\Delta_{2}} =
-\frac{C_{2}f_{5}v^{2}_{\phi}-C_{3}f_{2}v^{2}_{\zeta}}{(2f_{3}f_{5}-4f_{2}f_{6})v_{\Delta}}\frac{v^{2}_{\phi}}{\Lambda^{2}}
\end{eqnarray}
Obviously all the shifts in three and/or two components of all scalar fields in $\Phi^{e}_{1}$ are different but within the same order of magnitude.
All the shifts, however, in three and/or two components of all scalar fields in $\Phi^{\nu}$ are exactly the same, means the small shifts
are in the same direction of LO alignment. The result shows the stability of $\langle\Phi^{\nu}\rangle$, thus no soft terms are needed to drive the
superpotential into desired minimum.  The stable solutions of $\langle\Phi^{\nu}\rangle$ also guarantee the stability of TBM mixing in neutrino sector.
Take into account the conditions in Eq.~\eqref{eq:hierarchy VeVf}, it is easy to check the relative order of
shifted VEVs with respect to LO VEVs as following
\begin{eqnarray}\label{eq:dvorder}
\qquad\frac{\delta v_{\varphi_{i}}}{v_{\varphi}} \sim \frac{\delta v_{\eta_{1}}}{v_{\eta}} \sim \frac{\delta v_{\chi_{i}}}{v_{\chi}}
\sim\lambda^{4}_{c},
\quad \frac{\delta v_{\sigma}}{v_{\sigma}} \sim \lambda^{3}_{c},\quad
\frac{\delta v_{\xi_{2}}}{v_{\xi}} \sim \lambda^{2}_{c},\quad \frac{\delta v_{\xi_{1,3}}}{v_{\xi}}=0,\quad
\frac{\delta v_{\phi_{i}}}{v_{\phi}} \sim \frac{\delta v_{\Delta_{i}}}{v_{\Delta}} \sim \lambda^{6}_{c}
\end{eqnarray}
\end{appendix}

\begin{appendix}
\renewcommand{\theequation}{{C}\thesection\arabic{equation}}
\setcounter{equation}{0}
\section*{APPENDIX C: The Messenger sector}\label{C}
The higher dimensional operators of the effective superpotential can be obtained by integrating out the heavy messenger fields.
The messenger fields $\Gamma_{i}$ whose masses arise from singlets are listed in~\hyperref[ta:tab41]{Table 4}. Here only half of the messenger
fields are presented since each $\overline{\Gamma}_{i}$ takes opposite charges with respect $\Gamma_{i}$, which makes the direct mass term
$M_{\Gamma_{i}}\Gamma_{i}\overline{\Gamma}_{i}$ is guaranteed by the symmetries. The messenger pairs $\Sigma_{i}$ and $\overline{\Sigma}_{i}$
whose masses arise from the adjoint $H_{24}$ (of the form $H_{24}\Sigma_{i}\overline{\Sigma}_{i}$) are listed in~\hyperref[fig:T1H24]{Fig.~7}.
Due to the amount of the operators and the variety of their possible contraction ways of $S_{4}$, not all of the operators are listed.
Only those whose contributions to the entries of mass matrices non-vanishing at LO and small part of those being non-neglected at NLO
(or NNLO) corrections are showed below. 
Note that $\Gamma_{3}$ has all the $Z_{N}$ charges to be 1, thus we denote $\Gamma^{(r_{i})}_{3}$ as one symbol $\Gamma_{3}$ for simplicity.
In the left of~\hyperref[ta:tab5]{Table 5} $\Gamma_{3}$ contains all the possible $S_{4}$ representations, i.e., $1_{1}$, $2$ and $3_{1}$,
depend on the $S_{4}$ contraction ways of the fields in the corresponding operators. In the right table $\Gamma_{3}$ only include
the invariant $1_{1}$ of $S_{4}$. It is easy to realise the difference.

\begin{table}[!htbp]\footnotesize
  \centering
  \caption{Transformation properties of the heavy messenger fields with direct masses in the model.
  The superscript $r_{i}$ in $\Gamma^{(r_{i})}_{3}$ stands for the representations $1_{1}$, $2$ and $3_{1}$ of $S_{4}$.}\label{ta:tab41}
  \resizebox{\textwidth}{!}{
  \begin{tabular}{|ccccccccccccccccccccc|}
  \hline
  \hline
  Field & $\Gamma_{1}$ & $\Gamma_{2}$ & $\Gamma^{(r_{i})}_{3}$ & $\Gamma_{4}$ & $\Gamma_{5}$
  & $\Gamma_{6}$ & $\Gamma_{7}$ &$\Gamma_{8}$ & $\Gamma_{9}$ & $\Gamma_{10}$
  & $\Gamma_{11}$ & $\Gamma_{12}$ & $\Gamma_{13}$ & $\Gamma_{14}$ & $\Gamma_{15}$
  & $\Gamma_{16}$ & $\Gamma_{17}$ & $\Gamma_{18}$ & $\Gamma_{19}$ & $\Gamma_{20}$ \\
  \hline
  SU(5) & $\bar{\textbf{5}}$ & \textbf{10} & $\textbf{1}$ & $\bar{\textbf{5}}$ & \textbf{5}
  & \textbf{5} & $\bar{\textbf{5}}$ & $\bar{\textbf{5}}$ & $\bar{\textbf{5}}$ & $\bar{\textbf{5}}$
  & $\bar{\textbf{5}}$ & $\bar{\textbf{5}}$ & \textbf{10}& $\bar{\textbf{5}}$ & \textbf{1}
  & \textbf{10} & $\overline{\textbf{10}}$ & $\overline{\textbf{15}}$ & $\bar{\textbf{5}}$ & \textbf{1}\\
  $S_{4}$ & $3_{1}$ &  $1_{2}$ & $r_{i}$ & $1_{1}$ & $1_{2}$
  & $1_{2}$ & $1_{1}$ & $1_{2}$ & $1_{2}$ & $3_{2}$
  & $1_{1}$ & $1_{1}$ & $1_{1}$ & $1_{1}$¡¡& $1_{2}$
  & $3_{2}$ & $1_{2}$ & $1_{2}$ & $3_{2}$ & $3_{2}$ \\
  $Z_{4}$ & 1 & $-i$ & 1 & $-1$ & $i$
  & $-i$ & 1 & $-i$ & 1 & $-1$
  & $i$ & $-1$ & $-1$ & 1 & $-1$
  & 1 & $i$ & $-i$ & $-i$ & $i$ \\
  $Z_{6}$& $\omega^{2}$ & $-\omega$ & 1 & $-1$ & $-\omega^{2}$
  & $-\omega^{2}$ & $\omega$ & $-1$ & 1 & $\omega^{2}$
  & $\omega$ & $-\omega$ & $\omega^{2}$ & $\omega^{2}$ & $\omega^{2}$
  & $-\omega$ & $-\omega^{2}$ & $-\omega$ & $\omega^{2}$ & $\omega$ \\
  $Z_{5}$ & 1 & $\omega^{4}$ & 1 & $\omega$ & 1
  & 1 & 1 & $\omega$ & $\omega^{2}$ & 1
  & 1 & $\omega$ & $\omega^{3}$ & $\omega$ & $\omega$
  & $\omega^{2}$ & 1 & 1 & 1 & $\omega$  \\
  $Z_{2}$ & $-1$ & $-1$ & 1 & $-1$ & $-1$
  & $-1$ & 1 & $-1$ & 1 & 1
  & 1 & $-1$ & $-1$ & 1 & 1
  & 1 & $-1$ & $-1$ & $-1$ & 1   \\
  $U(1)_{R}$ & 0 & 0 & 0 & 0 & 0 & 0 & 0 & 0 & 0 & 0 & 0 & 0 & 0 & 0 & 0 & 0 & 0 & 0 & 0 & 0 \\
  \hline
  \hline
\end{tabular}}
\end{table}

\begin{figure}[!htb]
  \centering
  \subfigure[]{
  \label{fig:subfig:NF}
  \includegraphics[bb=70 565 535 720,scale=0.30]{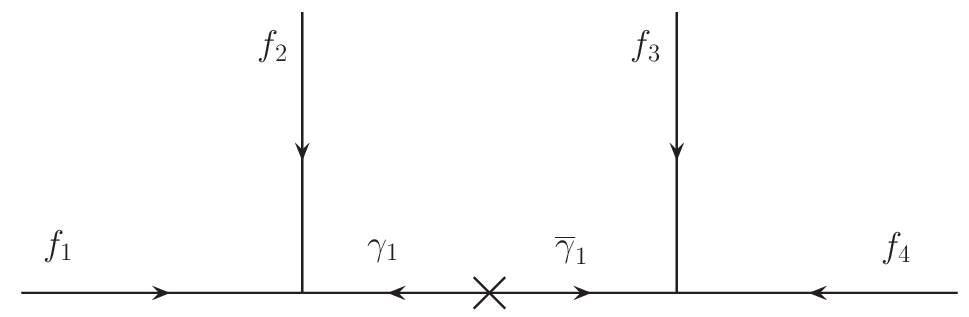}}
  \hspace{0.02in}
  \subfigure[]{
  \label{fig:subfig:T2T2}
  \includegraphics[bb=70 610 535 720,scale=0.40]{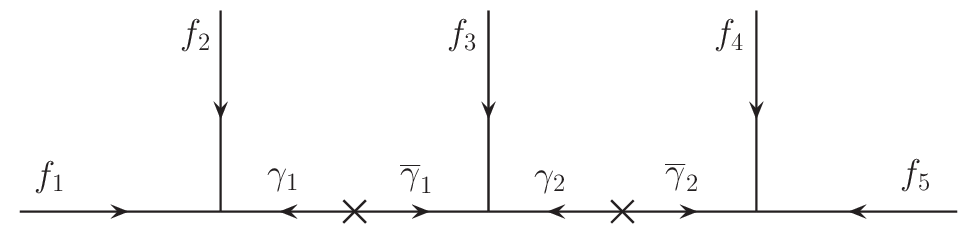}}
  \caption{The supergraphs before integrating out the messenger fields of order 4 (left) and 5 (right) operators in the superpotential.}
  \label{fig:SG1}
\end{figure}
\begin{table}[!htbp]\footnotesize
  \centering
  \caption{The operators corresponding to Fig.~\ref{fig:subfig:NF} (left) and Fig.~\ref{fig:subfig:T2T2}(right)}\label{ta:tab5}
  {
  \begin{tabular}{|c | c | c | c | c || c |}
  \hline
  Field & $f_{1}$ & $f_{2}$ & $f_{3}$ & $f_{4}$ & $\gamma_{1}$ \\
  \hline
  1 & $H_{5}$ & $N^{c}$ & F & $\phi/\Delta$ & $\Gamma_{1}$\\
  2 & $H_{45} $ & $T_{3}$ & $T_{2}$ & $\sigma$ & $\Gamma_{2}$\\
  3 & $N^{c}$ & $N^{c}$ & $\Phi^{\nu}$ & $\Phi^{\nu}$ & $\Gamma_{3}$\\
  4 & $T_{3}$ & $H_{\bar{5}}$ & F & $\varphi$ & $\Gamma_{4}$ \\
  \hline
  \end{tabular}
  \hspace{0.2in}
  \begin{tabular}{|c|c|c|c|c|c||c|c|}
    \hline
    fields & $f_{1}$ & $f_{2}$ & $f_{3}$ & $f_{4}$ & $f_{5}$ & $\gamma_{1}$ & $\gamma_{2}$ \\
    \hline
    1 & $T_{2}$ & $H_{5}$ & $T_{2}$ & $\sigma$ & $\vartheta$ & $\Gamma_{5}$ & $\Gamma_{6}$\\
    2 & $H_{5}$ & $T_{3}$ & $T_{3}$ & $\Phi^{\nu}_{i}$ & $\Phi^{\nu}_{i}$ & $\Gamma_{7}$ & $\Gamma^{1_{1}}_{3}$\\
    3 & $T_{2}$ & $H_{\overline{45}}$ & $\sigma/\xi$ & F & $\chi/\zeta$ & $\Gamma_{8}$ & $\Gamma_{9}/\Gamma_{10}$\\
    4 & $T_{1}$ & $H_{\overline{45}}$ & $\sigma$ & F & $\varphi$ & $\Gamma_{11}$ & $\Gamma_{12}$\\
    \hline
  \end{tabular}}
\end{table}
\begin{figure}[!htb]
  \centering
  \subfigure[]{
  \label{fig:subfig:T1T1}
  \includegraphics[bb=70 610 535 720,scale=0.40]{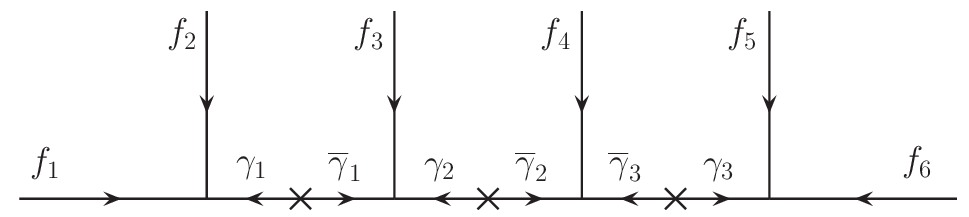}}
  \hspace{0.02in}
  \subfigure[]{
  \label{fig:subfig:T2H24}
  \includegraphics[bb=70 630 535 720,scale=0.49]{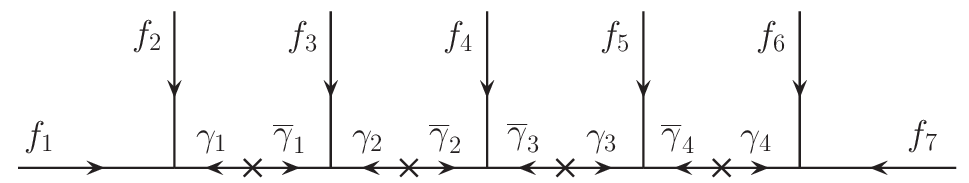}}
  \caption{The supergraphs of order 6 and 7 operators in the superpotential.}
  \label{fig:SG2}
\end{figure}
\begin{table}[!htbp]\footnotesize
  \centering
  \caption{The operators corresponding to Fig.~\ref{fig:subfig:T1T1}(up) and Fig.~\ref{fig:subfig:T2H24}(down)}\label{ta:tab6}
  {
  \begin{tabular}{|c | c | c | c | c | c | c || c | c | c |}
    \hline
    Field & $f_{1}$ & $f_{2}$ & $f_{3}$ & $f_{4}$ & $f_{5}$ & $f_{6}$ & $\gamma_{1}$ & $\gamma_{2}$ & $\gamma_{3}$ \\
    \hline
    1 & $T_{1}$ & $H_{5}$ & $T_{1}$ & $\zeta$ & $\eta$ & $\Delta$  & $\Gamma_{13}$ & $\Gamma_{14}$ & $\Gamma_{15}$\\
    2 & $H_{\overline{45}}$ & $T_{2}$ & $\sigma$ & F & $\varphi$ & $\eta$ &$\Gamma_{8}$ & $\Gamma_{9}$& $\Gamma_{16}$\\
    3 & $H_{45}$ & $T_{2}$ & $T_{3}$ & $\sigma$ & $\Phi^{\nu}_{i}$ & $\Phi^{\nu}_{i}$ &$\Gamma_{2}$ & $\overline{\Gamma}_{10}$ & $\Gamma^{1_{1}}_{3}$\\
    \hline
  \end{tabular}}
  \hspace{0.2in}
  {
  \begin{tabular}{|c|c|c|c|c|c|c|c||c|c|c|c|}
    \hline
    Field & $f_{1}$ & $f_{2}$ & $f_{3}$ & $f_{4}$ & $f_{5}$ & $f_{6}$ & $f_{7}$ & $\gamma_{1}$ & $\gamma_{2}$ & $\gamma_{3}$ & $\gamma_{4}$\\
    \hline
    1 & $\sigma$ & $T_{2}$ & $H_{24}$ & F & $H_{\bar{5}}$ & $\phi/\Delta$ & $\xi$ & $\Gamma_{17}$ & $\Gamma_{18}$ & $\Gamma_{19}$ & $\Gamma_{20}$ \\
    \hline
  \end{tabular}}
\end{table}

\begin{table}[!htbp]\footnotesize
\centering
\caption{The supergraph of operators with messenger masses from the adjoint $H_{24}$ and the assignments of GUT and flavor groups
of corresponding messenger fields.}{{
  \label{fig:T1H24}
  \includegraphics[bb=70 610 535 720,scale=0.50]{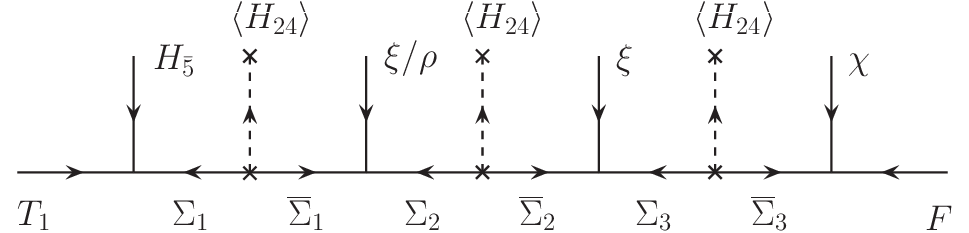}}
  \hspace{0.2in}
\begin{tabular}{|ccccccc|}
  \hline
  \hline
  Field & $\Sigma_{1}$ & $\overline{\Sigma}_{1}$ & $\Sigma_{2}$ & $\overline{\Sigma}_{2}$ & $\Sigma_{3}$ & $\overline{\Sigma}_{3}$ \\
  \hline
  SU(5) & $\bar{\textbf{5}}$ & \textbf{5} & $\bar{\textbf{5}}$ & \textbf{5} & $\bar{\textbf{5}}$ & \textbf{5} \\
  $S_{4}$ & $1_{1}$ &  $1_{1}$ & $3_{1}/1_{1}$ & $3_{1}/1_{1}$ & $3_{1}$ & $3_{1}$ \\
  $Z_{4}$ & 1 & $-1$ & $i$ & $i$ & $-1$ & 1 \\
  $Z_{6}$& $-\omega^{2}$ & 1 & $-\omega^{2}$ & 1 & $-\omega^{2}$ & 1  \\
  $Z_{5}$ & $\omega^{4}$ & $\omega$ & $\omega^{3}$ & $\omega^{2}$ & $\omega^{2}$ & $\omega^{3}$   \\
  $Z_{2}$ & 1 & 1 & $-1$ & $-1$ & 1 & 1   \\
  $U(1)_{R}$ & 0 & 0 & 0 & 0 & 0 & 0 \\
  \hline
  \hline
\end{tabular}}
\end{table}
\end{appendix}
\thispagestyle{empty}

\end{document}